\documentclass{aa}
\usepackage{graphicx}
\usepackage{savesym}
\usepackage{amsmath}
\savesymbol{iint}
\usepackage{txfonts}
\restoresymbol{TXF}{iint}
\usepackage{paralist}
\usepackage{wasysym}
\usepackage[normalem]{ulem}

\usepackage{natbib,twoopt}
\usepackage[breaklinks=true]{hyperref} 
\bibpunct{(}{)}{;}{a}{}{,}             
\makeatletter
  \newcommandtwoopt{\citeads}[3][][]{\href{http://adsabs.harvard.edu/abs/#3}%
    {\def\hyper@linkstart##1##2{}%
     \let\hyper@linkend\@empty\citealp[#1][#2]{#3}}}
  \newcommandtwoopt{\citepads}[3][][]{\href{http://adsabs.harvard.edu/abs/#3}%
    {\def\hyper@linkstart##1##2{}%
     \let\hyper@linkend\@empty\citep[#1][#2]{#3}}}
  \newcommandtwoopt{\citetads}[3][][]{\href{http://adsabs.harvard.edu/abs/#3}%
    {\def\hyper@linkstart##1##2{}%
     \let\hyper@linkend\@empty\citet[#1][#2]{#3}}}
  \newcommandtwoopt{\citeyearads}[3][][]%
    {\href{http://adsabs.harvard.edu/abs/#3}
    {\def\hyper@linkstart##1##2{}%
     \let\hyper@linkend\@empty\citeyear[#1][#2]{#3}}}
\makeatother

\begin{document}

\title{Fourier spectra from exoplanets with polar caps and ocean glint
}
\titlerunning{Fourier spectra from exoplanets with polar caps}
\author{P.M.\ Visser \and F.J.\ van de Bult}
\institute{
Delft Institute of Applied Mathematics, Technical University Delft,
Mekelweg 4, 2628 CD Delft, The Netherlands \\
\email{p.m.visser@tudelft.nl}
}
\date{Received 15 September 2014\ / Accepted 10 February 2015}
\abstract{The weak orbital-phase dependent reflection signal of an exoplanet contains information on the planet surface, such as the distribution of continents and oceans on terrestrial planets. This light curve is usually studied in the time domain, but because the signal from a stationary surface is (quasi)periodic, analysis of the Fourier series may provide an alternative, complementary approach.
}
{We study Fourier spectra from reflected light curves for geometrically simple configurations. Depending on its atmospheric properties, a rotating planet in the habitable zone could have circular polar ice caps.
Tidally locked planets, on the other hand, may have symmetric circular oceans facing the star.
These cases are interesting because the high-albedo contrast at the sharp edges of the ice-sheets and the glint from the host star in the ocean may produce recognizable light curves with orbital periodicity, which could also be interpreted in the Fourier domain.
}
{We derive a simple general expression for the Fourier coefficients of a quasiperiodic light curve in terms of the albedo map of a Lambertian planet surface.
Analytic expressions for light curves and their spectra are calculated for idealized situations, and dependence of spectral peaks on the key parameters inclination, obliquity, and cap size is studied.
}
{The ice-scattering and ocean glint contributions can be separated out, because the coefficients for glint are all positive, whereas ice sheets lead to even-numbered, higher harmonics. An in-view polar cap on a planet without axial tilt only produces  a single peak. The special situation of edge-on observation, which is important for planets in transit, leads to the most pronounced spectral behavior. Then the respective spectra from planets with a circumventing ocean, a circular ocean (eyeball world), polar caps, and rings, have characteristic power-law tails $n^{-2}$, $n^{-7/2}$, $n^{-4}$, and $(-1)^{n+1}n^{-2}$. 
}
{Promising recently discovered planetary systems may be selected as candidates for long-term (multiyear) observation: their Fourier spectra could separate the different planets and reveal or identify a water-covered planet with polar caps.
}
\maketitle

\begin{table*}[t]
\caption{Planet geometries and viewing configurations}
\label{table1}      
\centering
\resizebox{\hsize}{!}{
\includegraphics{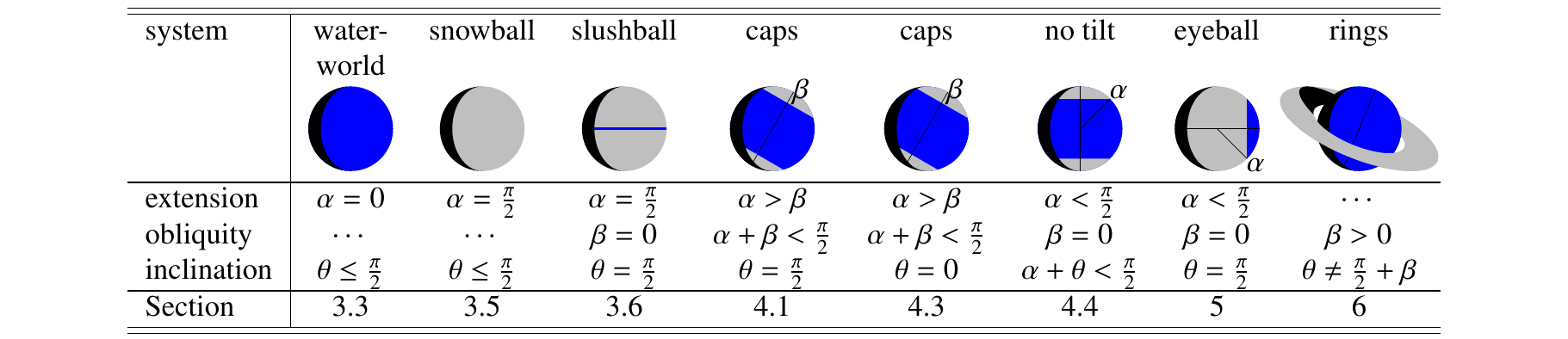}
}
\end{table*}

\section{Introduction}
In the past decade more than a thousand planets near other stars have been found, proving that nearly every star has planets. The methods of discovery range from direct detection, by watching their birth in the circumstellar planetary disks \citep{Greaves2005}, to more indirect detection in the stellar light dip during the passage of a transiting planet in front of its host star \citep{Hidas2005,Steffen2012} and in the time-varying phases from reflected star light \citep{Gaulme2010}. Indirectly inferring the planet's existence is possible by  monitoring  the motion of the host star \citep{Benedict2002} about the common center of mass via measurement of the star's position coordinates or its Doppler-shifted velocity \citep{Mayor1995}, or through  the detection of stellar light variations due to the effect of the ellipsoidal deformation  of the star by a planet \citep{Welsh2010}.

The phase light curve, which is the (for a twin Earth $10^{-10}$ times) weaker overall reflection signal of an orbiting planet on top of the direct stellar light, is a periodic function in time. This periodicity arises as the following combination of different effects: \begin{inparaenum}[(i)]
\item
For a noncircular orbit the distance to the star varies, so the planet generally scatters more light at periastron and less at  a-periastron (see Appendix).
\item
For inclined observation directions (not top view) the planet exhibits phases, creating  waxing and waning sickle shapes similar to the Moon's phases. A sickle shape gives off less light than the fully illuminated disk. Hence the overall (disk-integrated) intensity shows variations even though these shapes cannot be resolved \citep{Green2003,Dyudina2005,Snellen2009}.
\item 
For an inhomogeneous planet surface, the net reflection also changes with time because different regions are illuminated at different times as a consequence of rotation or orbital motion \citep{Ford2001}.
\end{inparaenum}
This phase-dependent light curve can, in fact, reveal part of the surface map \citep{Cowan2009,Cowan2013}. The study of a planet's surface requires capturing reflected photons, but not  spatially resolving the planet.

We advocate making an exceedingly long continuous observation of one exoplanetary system, as suggested by \citet{Cowan2008}, for the purpose of obtaining the Fourier transform of the net intensity.  Ideally, the observation should last several orbits and cover a large optical range, without a thermal component. It would be possible to use data from the Kepler satellite \citep{Borucki2009}, or a promising extrasolar system may be found by the future PLATO mission \citep{Catala2009}. The Fourier analysis of this long-duration campaign has clear advantages.
The bright glare from the parent star,
\begin{inparaenum}[(i)]
\item and
\item signal from irregular time-varying cloud patterns, and
\item random noise leave a continuous spectrum and can be filtered out (which is what Fourier transforms are generally used for). On the other hand, stationary surface patterns from land and oceans on the planet give spectral peaks (see the periodogram in \citep{Palle2008}. The diurnal (rotational) and annual (orbital) motion generally result in a mixed time signal, which could be messy, but
\item the spectrum has nonoverlapping peaks allowing spin-orbit tomography \citep[as proposed by][]{Fujii2012}.
Similarly, in a multiplanet system, orbital frequencies are usually incommensurable (not in orbital resonance) such that 
\item spectral peaks from different planets can also be separated out.
\end{inparaenum}
Signals originating from different effects can be distinguished more easily in the spectrum (vi). For example, diffusive (Lambertian) scattering only yields  even higher harmonics \citep[see][]{Cowan2013}, which could be used to separate it from other effects such as ocean-glint \citep{Williams2008,Oakley2009,Robinson2010,Zugger2010}.
The signal-to-noise ratio in spectral peaks of phase curves is enhanced, (vii) and may become comparable to a transit signal (after multiyear integration).

In this paper, we calculate the Fourier spectra of the light intensity from planets with particularly simple, but interesting geometries that may arise when the planet is covered only in water and ice. The geometry for a spinning planet can be described with two circular symmetric polar ice caps, or, when a planet is tidally locked, with a circular ocean facing the star. These cases are interesting because these planets fall right in the habitable zone, where water and ice coexist. The phase light curve for these geometries is strongest: Fourier peaks are most pronounced for these symmetric cases, and visibility should be optimal. We also exploit the property that liquid water predominantly reflects specularly and is otherwise a strong absorber, whereas snow is nature's best diffusive reflector. At the same time, a simple symmetric configuration leads to mathematical simplicity, with the possibility of deriving closed analytical expressions for key planetary parameters, such as polar cap extension $\alpha$ and obliquity $\beta$.

The structure of this paper is as follows. In Sect.\ \ref{SecII} we introduce the general framework of our model; the notation and our model assumptions. The influence of fluctuations from the stellar source and of surface-obscuring planetary clouds on the spectrum is briefly considered. In Sect.\ \ref{SecIII} we show that the scattering- and glint contributions $f_\mathrm{I}$ and $f_\mathrm{L}$ to the light curve are generally quasiperiodic, and then we demonstrate how the coefficients of the periodic spectrum of a banded planet are obtained. We derive exact formulas for these coefficients for the Lambertian phase curve and for the standard Fresnel light curve, corresponding to a snowball planet and an ocean-covered planet. In Sect.\ \ref{SecIV} we consider planets with polar caps, and derive and study the spectra for interesting special cases. The last two sections deal with the signals from a tidally-locked eyeball planet (Sect.\ \ref{SecV}) and planets with Saturn-like rings (Sect.\ \ref{SecVI}). These geometries are illustrated in Table \ref{table1}. In the appendices, we discuss how a transit affects the spectrum, address the problem of observation gaps, and show how the main results of the paper can be extended for noncircular planet orbits, by adding corrections due to the (small) ellipticity of a Kepler orbit.

\section{Fourier series for light curves}
\label{SecII}
The central star is placed at the origin of our coordinate system and the planet, with position $\boldsymbol r(t)$, moves in a circular orbit with time $t$ as
\begin{equation}
\boldsymbol r(t) = \hat{\boldsymbol r}(t)r = \boldsymbol ix(t) + \boldsymbol jy(t) = (\boldsymbol i\cos\omega t +\boldsymbol j\sin\omega t)r
.
\label{r}
\end{equation}
We denote with $\omega$ the (angular) orbital frequency: the orbital period is $2\pi/\omega$, one planet year. Influences from companion moons and other planets on $\boldsymbol r(t)$ are also neglected  (orbit eccentricity $\epsilon\neq 0$ is considered in Appendix \ref{AppB}). The polar axis $\hat{\boldsymbol n}$ of the planet lies in the $xz$ plane and has obliquity $\beta$. The observer at Earth is located (at a distance $L$) in the direction $\hat{\boldsymbol o}$:
\begin{align}
\hat{\boldsymbol n} &= \boldsymbol i\sin\beta + \boldsymbol k\cos\beta
, \label{n} \\
\hat{\boldsymbol o} &= (\boldsymbol i\cos\phi + \boldsymbol j\sin\phi)\sin\theta + \boldsymbol k\cos\theta
. \label{o}
\end{align}
It is natural to choose the obliquity (tilt angle) $\beta$ and the inclination (polar observation angle) $\theta$ in the interval $[0,\frac{\pi}{2}]$.

\subsection{Configurations and their symmetry}
We assume that the rotation frequency $\Omega$ is sufficiently slow that the planet is well approximated with a sphere \citep[for planet oblateness, see][]{Dyudina2005} and neglect precession on the timescale of observation. Ice sheets and land masses rotate about the axis $\hat{\boldsymbol n}$. In particular, polar caps do not grow and shrink significantly with the seasons. There are no effects from oceans waves or foam \citep{Vokroughlicky1995}. The atmosphere is at least partially transparent, such that seasonal clouds patterns contribute little, although irregular cloud patterns can lead to interference with the Fourier peaks. We will describe the surface with an (ice-) albedo map  and an (ocean-) reflectivity map ($M_\mathrm{I}$ and $M_\mathrm{L}$, respectively). Because the maps rotate about $\hat{\boldsymbol n}$, they have a periodic time dependence with period $2\pi/\Omega$. The annual and diurnal motion may give rise to periodicity in the reflected light signal. Of course, the diurnal period on the planet is $2\pi/(\Omega-\omega)$.

We neglected non-Lambertian scattering and polarization effects \citep{Stam2006,Zugger2010}. Thermal light from the planet is assumed to be absent or filtered out \citep[for the thermal light-signal, see][]{Cowan2012}. The star is also not deformed by the planet \citep[for ellipsoidal variations, see][]{Welsh2010}. We further make the standard approximations that the planet is much smaller than the central star $s\ll S$  ($s$ and $S$ are the respective radii of planet and star), the star is much smaller than orbital radius $S\ll r$, and relative motion of Earth (parallax) is irrelevant $r\ll L$. Since $s^2/4r^2$ is the fraction of star light received by the planet, we normalize the light curve $f(t)$ conveniently with $F(t)=f(t)Is^2/4r^2$,  with $I$, $F$ denoting the respective observed stellar and planet luminosities.

The reflected light curve for one planet is now in general a quasiperiodic function of time (derived in Sect.\ \ref{SecIII}), consisting of discrete peaks at multiples of the base frequencies $\omega$ and $\Omega$: \begin{equation}
f(t) = \sum_{n=0}^\infty \sum_{m=0}^\infty {\rm Re}\, (f^m_n \mathrm e^{\mathrm in\omega t+ \mathrm im\Omega t}) .
\label{quasiperiodic}
\end{equation}
Since $f$ is a linear function of the maps $M_\mathrm{I}$ and $M_\mathrm{L}$, and because the Fourier transform is linear, the Fourier coefficients $f^m_n$  are also linear functions of the planet surface maps.  \citet{Cowan2013} calculated the contributions from the spherical harmonics $Y^m_l(\vartheta,\varphi)$ to $f(t)$ (with respect to the rotation axis). Hence, there is also a linear transformation between their coefficients $C^m_l$ of the surface maps and the Fourier peaks $f^m_n$ defined here. The spherical wave expansion would be useful in our context; they may almost diagonalize the mapping.

For zero eccentricity, zero inclination, zero obliquity ($\epsilon=\theta=\beta=0$), a banded planet has no time-varying signal. A nonbanded planet for the same case, however, can have a diurnal signal: then $m$-peaks appear  only for $n=0$. If the vectors orbit-normal $\boldsymbol k$, the periastron $\boldsymbol i$ (in case of an ellipse), observation direction $\hat{\boldsymbol o}$, and polar axis $\hat{\boldsymbol n}$ (in case of a inhomogeneous surface) lie in one plane, then the light signal of a banded planet will be an even, periodic function with respect to the inferior conjunction, when the planet is in-between the star and Earth. There are $n$-peaks occurring only for $m=0$ and the ${f_n}^0$ are real numbers,
\begin{equation}
f(t) = \sum_{n=0}^\infty f^0_n \cos n\omega t .
\label{cosine}
\end{equation}
Otherwise, if these vectors break the orbital symmetry so that they are not in one plane, the light signal of a banded planet is not an even function, and contains cosine as well as sine components.

\subsection{Effect of stellar variability and cloud cover}
\label{SecIIB}
Because the reflected light signal involves the product of the source function $I(t)$ with $f(t)$, fluctuations in stellar intensity on orbital timescales disturb the signal. Consider, for example, an ideal light curve consisting of one harmonic at frequency $n\omega$. If we let $dI(t)$ represent the stellar noise, with variance $\mathrm{Var}\, I$, then the measured signal in the time domain is $F(t)=(I_0+dI(t))(f_0+f_n\cos n\omega t)s^2/4r^2$. The noise term $dI(t)f_0$ interferes with the wanted signal $I_0f_n\cos n\omega t$. One requires $I_0|f_n|\gg(\mathrm{Var}\, I)^{1/2}f_0$ to resolve the cosine. The noise for solar fluctuations at $10^{-8}\mathrm{Hz}$ is due to the $11$yr solar cycle, and has a level of $10^{-3}I_0$ \citep{Aigrain2004}. For a quiet solar-type star, the scattered light curve can be observed, in principle, for $|f_n|\gg 10^{-3}f_0$. See \citet{Lean2000}, for a reconstruction of the past $400 $yr solar irradiance. Of course, the photometry must then be sensitive enough to detect the fraction $f_n/f_0$ of the average planet luminosity. It is important to overcome the challenge of separating the direct starlight from the planet's reflection -- in the telescope, with a coronagraph or an occulting star shade -- especially if one wants to remove large stellar background noise too. If the starlight cannot be shielded, we require $s^2|f_n|\gg 10^{-3}r^2$. This is possible for Jupiter-sized planets near dwarf stars, as was done by \citet{Snellen2009}.

An astrophysicist, looking for planets around other stars, can search for peaks in the long duration intensity Fourier spectrum. In practice, the number of observable peaks in (\ref{quasiperiodic}) will be limited by stellar noise and observation time (and detector sensitivity). We  now estimate the noise-requirement for observing the peaks. For a finite observation time $T$, one may use the standard \emph{truncated\/} Fourier spectrum $F_T(\nu) = T^{-1/2}\int F(t)\mathrm e^{-\mathrm i\nu t}\mathrm dt$; the integration of $t$ is over $[0,T]$. For our periodic light curve (\ref{quasiperiodic}) this gives
\[
F_T(\nu) = \frac{s^2}{8r^2} \sum_{nm} [f^m_n I_T(\nu-n\omega-m\Omega) + (f^m_n)^* I_T(\nu+n\omega+m\Omega)]
.
\]
Hence, every peak becomes convoluted with the (truncated) spectrum $I_T(\nu)$ of the source $I(t)$. Noise at the frequency $\nu$ from the stellar source is quantified by the power spectrum $S_T(\nu)=|I_T(\nu)|^2$. For $T\longrightarrow\infty$, this function has a limit for all $\nu\neq 0$, with the lower frequencies having the slowest convergence. At $\nu=0$, $S_T$ has a peak of width $1/T$ and height $S_T(0)={I_0}^2T$. Thus, the central peak in $I_T(\nu)$ has strength $I_0T^{1/2}$. To observe the peak $f_n$ at $\nu=n\omega$ in $F(\nu),$ it is therefore required that $I_0 T^{1/2}|f_n|\gg S_T(n\omega)^{1/2}f_0$ (peak maximum above noise) and that $\omega T\gg 1$ (peak sufficiently narrow). We can now  estimate the noise level at $\nu=n\omega$,  using the property that the integrated power spectrum equals the variance $\mathrm{Var}\, I$. For a frequency interval of size $\Delta\nu>\omega$ around the peak at $n\omega$, one obtains $\mathrm{Var}\,I/\omega>\mathrm{Var}\,I/\Delta\nu>S_T(n\omega)$. Hence, if $\omega T>1$, the condition for the spectral noise may be more easily fulfilled than the corresponding variance requirement for a time series.
We need roughly $(\omega T)^{1/2}|f_n|\gg 10^{-3}f_0$. Hence, nine planet revolutions would enhance the signal with a factor $3$ over one round trip.

A similar analysis can be used to estimate the effect of irregular cloud patterns over a planet surface with, for example, (Titan-like) glinting lakes. If clouds obscure the surface $90\%$ of the time and expose surface lakes $10\%$ of the time, the effect is that the underlying periodic glint signal is multiplied with a fluctuating function with average $I_0=.1$ and variance $\mathrm{Var}\, I=.1-.1^2=.09$. The peaks appear for $(\omega T)^{1/2}|f_n|\gg (\mathrm{Var}\, I)^{1/2}f_0/I_0 = 3f_0$. This requires over $60$ periods. Of course, the clouds have their own (Lambertian) average contribution to the light curve.

If sharp distinct peaks are resolved in $F(\nu)$, the observer may obtain a base spectral structure with harmonics ${f_1}^0$, ${f_2}^0$, ${f_3}^0$  corresponding to the planet's annual motion. When side structures ${f_1}^1$, ${f_2}^1$, ${f_3}^1$, etc.\ of similar pattern can be resolved, this could indicate the rotation of a nonbanded planet (the planet has detectable spots). Ideally, the frequencies $n\omega+m\Omega$ have incommensurable base frequencies $\omega$, $\Omega$, because then none of the peaks interfere. The Fourier analysis of the reflection signal from exoplanets will obviously be spoiled if the star variability is strongly periodic. Likewise, surface mapping becomes impossible if the clouds patterns have diurnal periodicity themselves.

\section{Scattering and reflecting surfaces}
\label{SecIII}
We describe the planet surface with two maps, $M_\mathrm{I}$ and $M_\mathrm{L}$, for the high-albedo (ice, snow) regions and for oceans (liquid). We are interested in high-contrast situations
where $M_\mathrm{I}$ equals one on ice and zero elsewhere (expressions in the next section allow for an albedo factor). The ice scattering is assumed to be Lambertian, isotropically in all directions away from the surface. Ideally, the liquid map $M_\mathrm{L}$ also equals one or zero and indicates whether there is open ocean or not. (Reduced visibility from cloud cover could be accounted for with a reduction factor.) A planet with an ice and liquid  surface would give a phase light curve that is the sum of the two independent contributions
\[
f(t) = f_\mathrm{I}(t) + f_\mathrm{L}(t) ,
\]
where I is for ice (methane or water) and L is for liquid (methane or water).

A light ray from the central star that hits the planet surface has the (unit) direction vector $\hat{\boldsymbol r}$ and can be scattered in the direction $\hat{\boldsymbol o}$ of an observer on Earth. With $\boldsymbol s=\hat{\boldsymbol s}s,$ we  denote surface vectors (originating from the planet center). Only the surface segment that is the intersection of the illumination ${\cal S}'=\{\boldsymbol s\in{\cal S}|\boldsymbol s\bullet\boldsymbol r<0\}$ (day-side hemisphere) and the visibility ${\cal S}''=\{\boldsymbol s\in{\cal S}|\boldsymbol s\bullet\boldsymbol o>0\}$ (visible to the observer) contributes to $f$. The phase light curve, therefore, is a surface integral over the spherical \emph{lune\/} ${\cal S}'\cap{\cal S}''=\rightmoon$. For a diffusive scattering surface, the light curve contains two inner products, as in
\begin{equation}
f_\mathrm{I}(t) = \frac{4}{\pi} \int\int_{\rightmoon}
d^2\hat{\boldsymbol s}\, ({-\hat{\boldsymbol r}(t)}\bullet\hat{\boldsymbol s})(\hat{\boldsymbol s}\bullet\hat{\boldsymbol o})M_\mathrm{I}(\hat{\boldsymbol s},t)
.
\label{formfactor}
\end{equation}
The explicit time-dependence in the albedo map $M_\mathrm{I}(\hat{\boldsymbol s},t)$ accounts for the rotation. The intensity is properly normalized: using
$
\iint_{\cal S''} d^2\hat{\boldsymbol o}\, (\hat{\boldsymbol s}\bullet\hat{\boldsymbol o}) = \pi
$, $
\iint_{\cal S'} d^2\hat{\boldsymbol s}\, ({-\hat{\boldsymbol r}}\bullet\hat{\boldsymbol s}) = \pi
$,
the overall scattering for a planet with unity albedo $M=1$ is precisely $\iint_{\cal S} d^2\hat{\boldsymbol o}\, f_\mathrm{I}(t)=4\pi$. The surface $\cal S$ is parametrized with the vector function $\boldsymbol s(\mu,\nu)$ and the surface element in the above integrals needs to be calculated with the Jacobian
\begin{equation}
d^2\hat{\boldsymbol s} = \Big|\frac{\partial\hat{\boldsymbol s}}{\partial\mu}\times\frac{\partial\hat{\boldsymbol s}}{\partial\nu}\Big| d\mu\, \mathrm d\nu
.
\label{element}
\end{equation}
We use $M_\mathrm{I}(\hat{\boldsymbol s},t)=M_\mathrm{I}(\mu,\nu,t)$ for rotating maps (ice, land, ocean, or otherwise) in non-rotating spherical coordinates $(\mu,\nu)$; see Table \ref{table2} for the parameters used. To find the corresponding coordinates $(\vartheta,\varphi)$ on the fixed map $M_\mathrm{I}(\vartheta,\varphi)$ in the rotating frame with respect to the axis $\hat{\boldsymbol n}$ as function of $(\mu,\nu,t)$, one may use the following relation:
$\hat{\boldsymbol s}\bullet(\boldsymbol j\times\hat{\boldsymbol n}+\mathrm i\boldsymbol j)=$
\[
\mathrm e^{\mathrm i\varphi+\mathrm i\Omega t}\sin\vartheta = \cos\nu\sin\mu\cos\beta-\cos\mu\sin\beta+\mathrm i\sin\nu\sin\mu
.
\]

\begin{table}
\caption{Parameters used in modeling}
\label{table2}      
\centering
\begin{tabular}{l|l}
\hline
\hline
symbol & quantity \\
\hline
$t$ & time \\
$2\tau$ & transit duration, or gap duration \\
$T$ & observation duration \\
$I(t)$ & star luminosity, power output \\
$I_T(\nu)$ & Fourier transform of star luminosity \\
$I_0$ & star luminosity average \\
$S(\nu)$ & power spectrum of $I(t)$ \\
$L$ & distance from Earth \\
$S$ & star radius \\
$s$ & planet radius \\
$\boldsymbol s=\hat{\boldsymbol s}s$ & planet surface vector \\
$r$ & planet-star distance \\
$\boldsymbol r=\hat{\boldsymbol r}r$ & planet position vector \\
$a$ &
orbit semimajor axis \\
$\epsilon$ &
orbit eccentricity \\
$\omega$ & orbital angular frequency, mean motion \\
$\omega t$ & orbital phase angle, mean anomaly \\
$\nu$ & frequency variable; true anomaly (in App. \ref{AppB}) \\
$\lambda$ & eccentric anomaly \\
$\Omega$ & rotational/spin angular frequency \\
$\hat{\boldsymbol n}$ & rotation/spin axis \\
$\beta$ & obliquity, tilt angle of rotation axis \\
$\alpha$ & polar cap extension angle from north pole \\
$b$, $c$ & inner ring radius, outer ring radius \\ 
${\cal T}$, ${\cal R}$ & ring transmitivity, ring- or ocean reflectivity \\
$M_\mathrm{I}$ & \emph{ice\/} cap/continent (albedo) map \\
$M_\mathrm{L}$ & \emph{liquid\/} lake/ocean surface map \\
\hline
$(x,y,z)$ & various rectangular coordinates \\
$(\rho,\nu,z)$ & cylindrical coordinates in inertial frame $\boldsymbol i$, $\boldsymbol j$, $\boldsymbol k$ \\
$(s,\mu,\nu)$ & radius, polar and azimuth angle in inertial frame \\
$(s,\vartheta,\varphi)$ & radius, polar and azimuth angle in rotating frame \\
\hline
$\theta$ & inclination, (polar) observation angle \\
$\theta=0$ & observation is face-on (orbital plane) \\
$\theta=\frac{\pi}{2}$ & observation is edge-on (orbital plane) \\
$\phi$ & azimuthal observation angle \\
$\omega t=\phi$ & phase at inferior conjunction \\
$F(t)$ & planet luminosity \\
$F_T(\nu)$ & Fourier transform of planet luminosity \\
$f(t)$ & light curve, normalized to $1$ for a mirror ball \\
$f_n^m$ & Fourier coefficient for light curve \\
$g(\nu,t)$ & pattern function, planet map integrated over $z$ \\
$g_n^m$ & Fourier coefficient for pattern function \\
\hline
\hline
\end{tabular}
\end{table}

\subsection{Fourier spectrum from Lambertian surface maps}
We are now ready to derive a general expression relating spectral peaks of the signal (\ref{formfactor}) to a surface albedo map $M_\mathrm{I}$. For a unit vector $\hat{\boldsymbol s}$ on the unit sphere in cylindrical coordinates $(\rho,\nu,z)$, we substitute $\rho=\sqrt{1-z^2}$ so that the planet surface is parametrized with $(\nu,z),$ as
\[
\hat{\boldsymbol s} = (\boldsymbol i\cos\nu+\boldsymbol j\sin\nu)\sqrt{1-z^2} + \boldsymbol k z
.
\]
It follows that
\begin{align*}
-\hat{\boldsymbol s}\bullet\hat{\boldsymbol r} &= -\sqrt{1-z^2}\cos(\omega t-\nu)
, \\
\hat{\boldsymbol s}\bullet\hat{\boldsymbol o} &= \sqrt{1-z^2}\sin\theta\cos(\nu-\phi) + z\cos\theta
.
\end{align*}
The circle that borders the visible region is found for $\hat{\boldsymbol s}\bullet\hat{\boldsymbol o}=0$, hence the limb has the coordinate
\begin{equation}
z_\mathrm{limb}(\nu) = \frac{-\cos(\nu-\phi)}{\sqrt{\cot^2\theta+\cos^2(\nu-\phi)}}
. \label{limb}
\end{equation}
Using the inner products and the surface element $d^2\hat{\boldsymbol s}=dz\, \mathrm d\nu$, the double integral in (\ref{formfactor}) is expressed as the following repeated integral:
\[
f_\mathrm{I}(t) =
\frac{4}{\pi}\!\int\limits_{\omega t-\frac{3\pi}{2}}^{\omega t-\frac{\pi}{2}}\! \mathrm d\nu\, |\cos(\omega t-\nu)| \frac{g(\nu,t)}{2}
,
\]
where the \emph{pattern function\/} $g(\nu,t)$ contains the $z$ integration over the interval between the limb point (\ref{limb}) and the point with $z=1$, i.e.,
\begin{equation}
\frac{g(\nu,t)}{2} = \int\limits_{z_\mathrm{limb}(\nu)}^1 \mathrm dz\, (\hat{\boldsymbol s}\bullet\hat{\boldsymbol o})\sqrt{1-z^2} M_\mathrm{I}(\hat{\boldsymbol s},t)
\label{g}
.
\end{equation}
Since the rotation axis $\hat{\boldsymbol n}$ and the observation direction $\hat{\boldsymbol o}$ are constant, $g(\nu,t)$ characterizes the scattering pattern from the visible regions. When $\nu$ and $t$ are considered independent variables, then $M_\mathrm{I}(\hat{\boldsymbol s}(\nu,z),t)$ is periodic in $\nu$ (because $\hat{\boldsymbol s}$ is periodic in $\nu$) and  in $\Omega t$ (modulo $2\pi$) and therefore
\begin{equation}
g(\nu,t) =
\sum_{n=0}^\infty \sum_{m=0}^\infty {\rm Re}\Big( \mathrm e^{\mathrm in\nu+\mathrm im\Omega t} g^m_n \Big)
.
\label{gn}
\end{equation}
When we make the substitution $\nu=\omega t+\bar\nu$
\[
\int\limits_{\omega t-\frac{3\pi}{2}}^{\omega t-\frac{\pi}{2}}\mathrm d\nu\, |\cos(\omega t-\nu)|\, \mathrm e^{\mathrm in\nu} = \mathrm e^{\mathrm in\omega t} \int\limits_{-\frac{3\pi}{2}}^{-\frac{\pi}{2}}\mathrm d\bar\nu\, |\cos\bar\nu|\, \cos n\bar\nu
,
\]
and use this relation in the remaining integral to get
\[
f_\mathrm{I}(t) = \frac{4}{\pi} \sum_{n=0}^\infty \sum_{m=0}^\infty {\rm Re}\Big( \mathrm e^{\mathrm in\omega t+\mathrm im\Omega t}\, \frac{g^m_n}{2}\Big) \int\limits_{-\frac{3\pi}{2}}^{-\frac{\pi}{2}}\mathrm d\bar\nu\, |\cos\bar\nu|\, \cos n\bar\nu
.
\]
We find that the Fourier coefficients of $f_\mathrm{I}$ are each related to the corresponding Fourier coefficients of $g$ by
\begin{equation}
f^m_0 = \frac{4}{\pi} g^m_0 , \quad
f^m_1 = -g^m_1 ,  \quad
f^m_{2n} = \frac{-(-1)^n}{4n^2-1} \frac{4}{\pi}g^m_{2n} , \quad
f^m_{2n+1} = 0
.
\label{fg}
\end{equation}
The expressions (\ref{fg}) are  key results of this paper. The Fourier coefficients $g^m_n$ of the pattern function $g(\nu,t)$ found from the surface map $M_\mathrm{I}(\hat{\boldsymbol s},t)$ immediately gives us the spectrum of the quasiperiodic light curve of a planet.

There are several reasons why the linear transformation from $M_\mathrm{I}$ to $f^m_n$ is not invertible.
Firstly, the region with polar angles $\vartheta>\frac{\pi}{2}+\arccos(\hat{\boldsymbol n}\bullet\hat{\boldsymbol o})$ below the southern circle (i.e.,\ $(\hat{\boldsymbol n}\bullet\hat{\boldsymbol o})^2+(\hat{\boldsymbol n}\bullet\hat{\boldsymbol s})^2<1$) is permanently inaccessible to the observer. Thus any map that is zero outside and arbitrary inside this circle gives no signal. \citep[These maps are in the null space of the transformation, see][]{Cowan2013}
Secondly, consider the case that $M_\mathrm{I}$ is only a function of $z$, not of $\nu$ and $t$. This is the case for a banded planet with zero obliquity (see \ref{SecIVC}). If a banded map is also zero below the northern circle $z=\sin\theta$, it can be seen that the integral in expression (\ref{g}) for $g(\nu)$ effectively runs over $z$-values from $\sin\theta$ to $1$. The dependence on $\nu$ comes from the factor $(\hat{\boldsymbol s}\bullet\hat{\boldsymbol o})$ only, which yields a single cosine. Therefore, the spectrum has just two peaks, with amplitudes ${f_0}^0$ and ${f_1}^0$, and most of the information from the banded structure is lost.
Thirdly, if a banded planet with zero obliquity is also observed edge-on, the inclination is $\theta=\frac{\pi}{2}$ and $z_\mathrm{limb}=\pm 1$. Then the $z$ integral vanishes for $+1$ and gives the same overall albedo constant for $-1$. This implies that the light curve is indistinguishable from that of a homogeneous planet.
Finally, if the rotation period and orbital period are commensurable, which is when $\Omega/\omega$ is a rational number, then different combinations of the coefficients $f^m_n$ have the same harmonic frequency and thus interfere. This is the case in particular if the ratio is not very large (or very small). Different maps can give the same signal, and detailed structure is lost.

\subsection{Fourier coefficients for a banded planet}
The surface pattern of a rotating planet is stationary only if the pattern is banded; that is, if the rotation axis is also an axis of symmetry for the pattern. In other words, the planet has no spots (like Jupiter or Neptune). The function $M_\mathrm{I}(\hat{\boldsymbol s},t)$ is invariant under rotation about $\hat{\boldsymbol n}$ and constant in time; then the planet's rotation is unobservable. It follows from (\ref{g}-\ref{gn}) that $g^m_n=0$ for all $m$ except $m=0$. In what follows, we only consider banded planets (and tidally locked planets, with $\Omega=\omega$), and, therefore, drop the index $m$ from all following expressions.

If we assume the further (reflection) symmetry $M_\mathrm{I}(-\hat{\boldsymbol s})=M_\mathrm{I}(\hat{\boldsymbol s})$, we can simplify the expression for the coefficients $g_n$. Let us consider first $g_1$. We make use of the fact that $\mathrm e^{-\mathrm i\nu}$ is opposite for values of $\nu$ that are $\pi$ apart. Points with $z>0,$  which are invisible, are opposite to points with $z<0$, which  are visible. This allows the $z$ integral (\ref{g}) for $g(\nu)$ to be split into a sum of integrals for two hemispheres, and to condense into one $\rho$ integral over the north hemisphere. Using $\rho dz=-(\rho^2/z)d\rho$, we may write
\begin{equation}
g_1 = \frac{2}{\pi} \int\limits_{-\pi}^\pi \mathrm d\nu\, \mathrm e^{-\mathrm i\nu}\int\limits_0^1 \mathrm d\rho\, \frac{(\hat{\boldsymbol s}\bullet\hat{\boldsymbol o})\rho^2}{\sqrt{1-\rho^2}} M_\mathrm{I}(\hat{\boldsymbol s})
. \label{g1}
\end{equation}
For even coefficients, the phase-factor $\mathrm e^{-2in\nu}$ for opposite points is the same, hence
\begin{equation}
g_{2n} = \frac{2}{\pi} \int\limits_{-\pi}^\pi \mathrm d\nu\, \mathrm e^{-2\mathrm in\nu}\int\limits_0^1 \mathrm d\rho\, \frac{|\hat{\boldsymbol s}\bullet\hat{\boldsymbol o}|\rho^2}{\sqrt{1-\rho^2}} M_\mathrm{I}(\hat{\boldsymbol s}) , \quad n\geq 1
. \label{g2n}
\end{equation}
The dependence on limb position in (\ref{g}) has been removed in (\ref{g1}-\ref{g2n}). Note the distinct absolute-value of the inner product in (\ref{g2n}) is absent in (\ref{g1}). The assumption of reflection symmetry must be independently validated before these expressions may be used.

\subsection{Fourier coefficients for a homogenous planet}
Consider the case of a planet with a homogeneous scattering surface. The orientation of the rotation axis is irrelevant and the symmetry about the $z$ axis is only broken by the fact that the observer is viewing from an inclined direction, and we may set $\phi=0$ in (\ref{o}). We calculate  the light curve $f(t)$ and its Fourier coefficients $f_n$. It turns out that, on the one hand, the time-signal $f(t)$ is most conveniently evaluated using a new coordinate system that exploits the symmetry about $\hat{\boldsymbol o}$, but that the peak intensities $f_n$, on the other hand, are most directly derived from the function $g(\nu)$. We thus write the surface vector $\boldsymbol s$ in spherical coordinates $(s,\mu,\nu)$ with respect to an orthogonal basis fixed by $\boldsymbol o$ and $\boldsymbol r$ as
\[
\hat{\boldsymbol s} = \frac{(\hat{\boldsymbol r}\times\hat{\boldsymbol o})\times\hat{\boldsymbol o}}{|\hat{\boldsymbol r}\times\hat{\boldsymbol o}|} \cos\nu\sin\mu + \hat{\boldsymbol o} \sin\nu\sin\mu + \frac{\hat{\boldsymbol r}\times\hat{\boldsymbol o}}{|\hat{\boldsymbol r}\times\hat{\boldsymbol o}|} \cos\mu
.
\]
We require for (\ref{formfactor}) the inner products
\begin{align*}
{-\hat{\boldsymbol s}}\bullet\hat{\boldsymbol r} &= \Big( \sqrt{1-(\hat{\boldsymbol r}\bullet\hat{\boldsymbol o})^2} \cos\nu - \hat{\boldsymbol r}\bullet\hat{\boldsymbol o}\sin\nu \Big) \sin\mu
, \\
\hat{\boldsymbol s}\bullet\hat{\boldsymbol o} &= \sin\nu\sin\mu
,
\end{align*}
and the surface element (\ref{element}) is $\mathrm d^2\hat{\boldsymbol s}=\sin\mu\, \mathrm d\mu\, \mathrm \mathrm d\nu$. Let $\varphi$ be the scattering angle, between $\boldsymbol r$ and $\boldsymbol o$, so that
\[
\cos\varphi = \hat{\boldsymbol r}\bullet\hat{\boldsymbol o} = \sin\theta\cos\omega t
.
\]
The condition $\hat{\boldsymbol s}\bullet\hat{\boldsymbol o}>0$ implies that $\mu$ lies in the interval $[0,\pi]$ and $\hat{\boldsymbol r}\bullet\hat{\boldsymbol s}<0$ implies that $\nu$ lies in the interval $[0,\varphi]$. With these definitions, and by setting $M_\mathrm{I}=1$, formula (\ref{formfactor}) reduces to a well-known form \citep{vanHulst1980}
\[
f_\mathrm{I} = \frac{4}{\pi}\int\limits_0^\pi \mathrm d\mu\int\limits_0^\varphi \mathrm d\nu\, \sin\mu\, (-\hat{\boldsymbol r}\bullet\hat{\boldsymbol s})(\hat{\boldsymbol s}\bullet\hat{\boldsymbol o})
= \frac{8\sin\varphi-8\varphi\cos\varphi}{3\pi}
.
\]
Completely expanded, this yields
\[
f_\mathrm{I}(t)=
\frac{8}{3\pi}\Bigg( \sqrt{1-\sin^2\theta\cos^2\omega t} - \sin\theta\cos\omega t \cos^{-1}(\sin\theta\cos\omega t) \Bigg)
.
\]
Note that, for top view $\theta=0$, the signal is constant. As previously indicated, for the Fourier components, we substitute $M_\mathrm{I}=1$, $\phi=0$ and (\ref{limb}) in the pattern function (\ref{g}) and obtain
\[
g(\nu) = \frac{4}{3} \Bigg( \sqrt{1-\sin^2\theta\sin^2\nu} - \frac{\cos^2\theta}{2\sqrt{1-\sin^2\theta\sin^2\nu}} +  \sin\theta\cos\nu  \Bigg)
.
\]
There are two steps to obtain  the Fourier coefficient from this function. First, calculate the Taylor series in the powers $x^{2k}=\sin^{2k}\theta$, and, then, expand
\[
\sin^{2k}\nu = \binom{2k}{k} \frac{1}{2^{2k}} + \sum_{n=1}^k \binom{2k}{k-n} \frac{(-1)^n}{2^{2k-1}}\cos 2n\nu
.
\]
Collecting the coefficients in front of the cosines leads to the result
\[
g_0 = \frac{8E-4K\cos^2\theta}{3\pi} , \quad g_1 = \frac{4\sin\theta}{3} ,  \quad g_{2n+1} = 0 .
\]
The functions $K=K(\sin\theta)$ and $E=E(\sin\theta)$ are the complete elliptic integrals of the first and second kind; the inclination $\theta$ has the role of the so-called modular angle. The higher-order coefficients lead to the hypergeometric function $_2F_1(a,b;c,x)$:
\begin{align*}
g_{2n} &= \frac{8\sin^{2n}\theta}{3\ 2^{2n}} \binom{\tfrac{1}{2}}{n}\  _2F_1(n+\tfrac{1}{2},n-\tfrac{1}{2} ; 2n+1 ; \sin^2\theta) \\
& - \frac{4\sin^{2n}\theta \cos^2\theta}{3\ 2^{2n}} \binom{-\tfrac{1}{2}}{n}\  _2F_1(n+\tfrac{1}{2},n+\tfrac{1}{2} ;2n+1 ; \sin^2\theta) .
\end{align*}
The hypergeometric functions in $g_{2n}$ can be expanded into $K$ and $E$  . With (\ref{fg}), the peak values $f_n$ follow. We obtain for the first four coefficients
\begin{align}
f_0 &= \frac{32E-16K\cos^2\theta}{3\pi^2} , \quad f_1 = -\frac{4\sin\theta}{3} ,
\label{fsnowtheta} \\
f_2 &= \frac{32K(2-3\sin^2\theta)\cos^2\theta-64E\cos 2\theta}{27\pi^2\sin^2\theta} , \quad f_3 = 0 .
\nonumber
\end{align}

\begin{figure}
\centering
\resizebox{\hsize}{!}{
\includegraphics{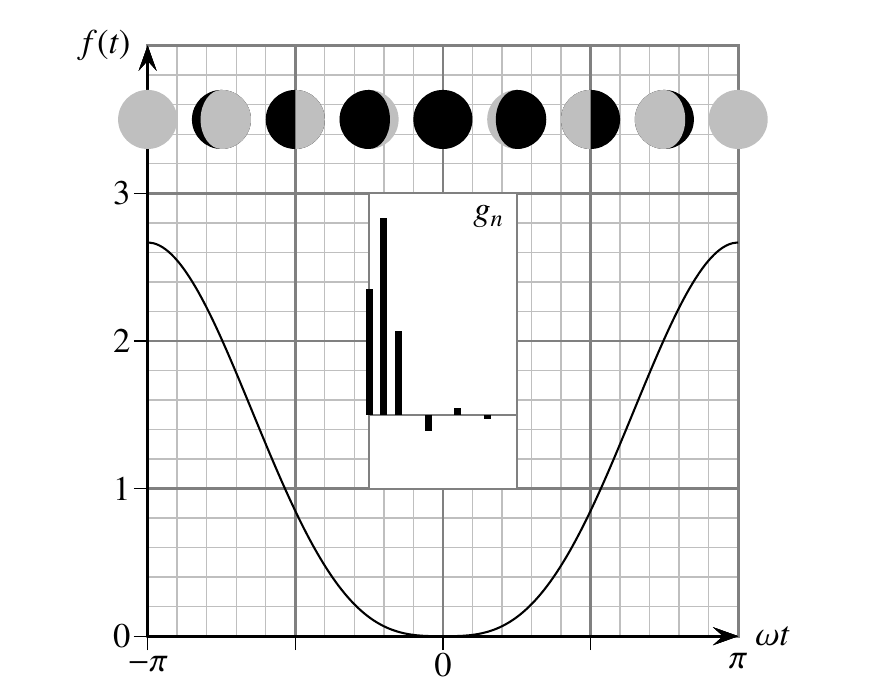}
}
\caption{\label{snowball}
Standard light curve (\ref{snow}) for a homogeneous (Lambertian) planet: normalized intensity $f$ versus orbital phase $\omega t$, for edge-on observation. The inset shows, to scale, the coefficients $g_n$ which are proportional to $f_n$ (\ref{fsnow}) according to (\ref{fg}).}
\end{figure}

\begin{figure}
\centering
\resizebox{\hsize}{!}{
\includegraphics{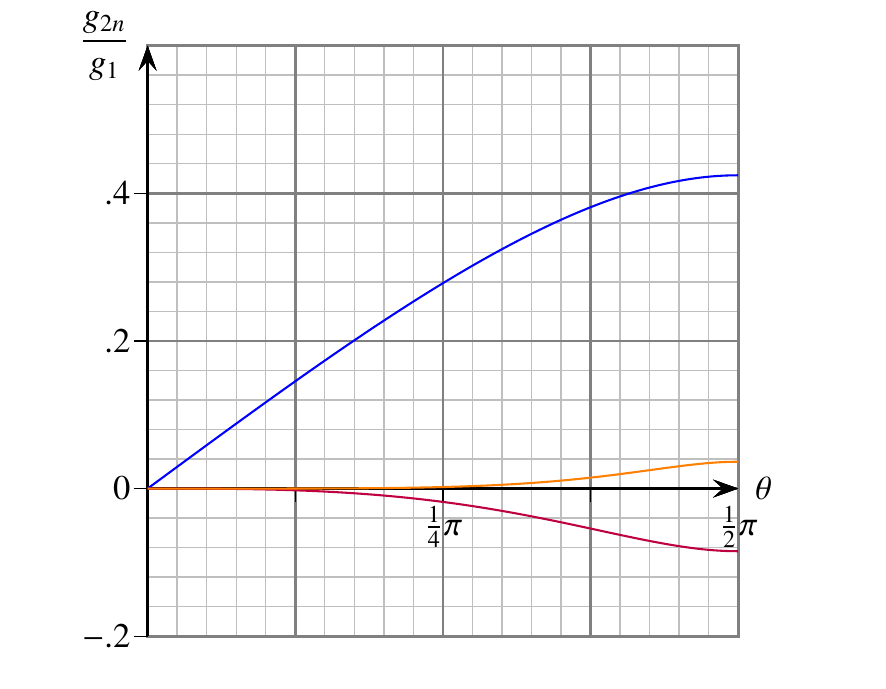}
}
\caption{\label{f2f1}
Harmonic ratios $g_2/g_1$ (blue), $g_4/g_1$ (purple) and $g_6/g_1$ (orange) for a snowball planet, versus inclination angle $\theta$. These curves can be used to determine the inclination $\theta$ or to conclude that the planet is, instead, inhomogeneous.
}
\end{figure}

If one observes the orbital plane edge on, then $\theta=\frac{\pi}{2}$ and $\varphi(t)=|\omega t|$ for $-\pi\leq\omega t\leq\pi$. The standard Lambertian phase curve for in-plane observation of a planet or moon is
\begin{equation}
f_\mathrm{I}(t) = \frac{8\sin|\omega t|-8|\omega t|\cos\omega t}{3\pi} , \quad -\pi\leq\omega t\leq\pi
\label{snow}
,\end{equation}
with Fourier coefficients,
\begin{equation}
f_0 = \frac{32}{3\pi^2} , \quad 
f_1 = \frac{-4}{3} , \quad f_{2n} = \frac{64}{3\pi^2}\, \frac{1}{(4n^2-1)^2}
.
\label{fsnow}
\end{equation}
The graph is plotted in Fig.\ \ref{snowball}. The coefficients (\ref{fsnowtheta}), (\ref{fsnow}) drop to zero with the fourth power in $n$, such that $f_4$ is already much smaller than $|f_1|$. Fig.\ \ref{f2f1} shows the ratios of coefficients $g_n$ as function of observation angle $\theta$. If  the planet radius $s$ and albedo were determined with other methods, then the first coefficient $f_1$ could be used to estimate or verify obliquity. Otherwise, the relative values of the peaks in the spectrum must be used to obtain orbital parameters. For example, the ratio of the first two peaks $f_2/f_1=-4g_2/3\pi g_1$ could be used to find $\theta$. For this purpose, in Fig.\
\ref{f2f1} we  plotted  the ratios as a function of $\theta$. The overall complex phase $\phi$ of the peaks, which was set to zero for the sake of convenience in our calculations, is just the 
orbital phase with respect to the inferior conjunction (in-between Earth and the star). The absence of $f_3$, $f_5$ would indicate that the planet is a diffusive scatterer without specular glint. The condition $|f_n|\gg 10^{-3}f_0$, derived in Sec.\ \ref{SecIIB}, valid for a sun-type star observed for one year, implies that $f_8$ is swamped in the stellar noise.

\subsection{Glint spot on a specular reflecting surface}
Consider a reflecting sphere with radius $s$, placed at the origin, which is illuminated by a point source on the $z$-axis at $z=+\infty$. Imagine a long cylinder of parallel light rays around the $z$ axis with radius $\rho<s$ and cross-section $A=\pi\rho^2$. The rays inside this cylinder hit the north pole of our sphere from angle zero up to angle $\theta$. Since $s$ is the radius of the sphere, this implies $\rho=s\sin\theta$. At the border, the angle of incidence is also $\theta$, so the cylindrical beam is reflected into a segment of the hemisphere with limiting angle $2\theta$. The total solid angle of the reflected light is therefore given by the integral $\int_0^{2\pi}\mathrm d\varphi\int_0^{2\theta}\mathrm d\vartheta\sin\vartheta=4A/s^2$. Because this solid angle is proportional to $A$, a perfectly reflecting sphere scatters light isotropically in all directions. Since a finite source (at large distance) may be composed of many point sources, a reflecting sphere also scatters the light of a finite source isotropically. The reflection of a metallic planet would, like a garden ball, be equally bright from all directions, and there would be no variation in intensity during its yearly orbit.

If the light source is a star with radius $S$ and our sphere is a planet with radius $s$, the light rays that are incident on the smaller planet are confined to the cone $r\hat{\boldsymbol r}+r\mathrm d\hat{\boldsymbol r}$. Here $\mathrm d\hat{\boldsymbol r}$ is orthogonal to $\hat{\boldsymbol r}$ and its length is $|\mathrm d\hat{\boldsymbol r}|=S/r$. Since the distance to the observer is very large, the reflected cone is practically collapsed into a line with direction $\boldsymbol o=L\hat{\boldsymbol o}$. The points where the light rays are reflected on the surface, form a small ellipse around the central point of reflection
\[
\boldsymbol{\hat s} = \frac{\hat{\boldsymbol o}-\hat{\boldsymbol r}}{|\hat{\boldsymbol o}-\hat{\boldsymbol r}|} = \frac{\hat{\boldsymbol o}-\hat{\boldsymbol r}}{2\cos\vartheta}
.
\]
Here $\vartheta$ is the angle of incidence with respect to the surface normal, with $\cos\vartheta=-\hat{\boldsymbol s}\bullet\hat{\boldsymbol r}\geq 0$. This ellipse is now given by
\[
\hat{\boldsymbol s} + \mathrm d\hat{\boldsymbol s} = \frac{\hat{\boldsymbol o}-\hat{\boldsymbol r}-\mathrm d\hat{\boldsymbol r}}{|\hat{\boldsymbol o}-\hat{\boldsymbol r}-\mathrm d\hat{\boldsymbol r}|} = \frac{\hat{\boldsymbol o}-\hat{\boldsymbol r}}{|\hat{\boldsymbol o}-\hat{\boldsymbol r}|} + \frac{(\hat{\boldsymbol s}\bullet \mathrm d\hat{\boldsymbol r})\hat{\boldsymbol s}-\mathrm d\hat{\boldsymbol r}}{2(-\hat{\boldsymbol s}\bullet\hat{\boldsymbol r})}
.
\]
The vector $\mathrm d\hat{\boldsymbol s}$ points from the center to the edge of the ellipse and is orthogonal to $\hat{\boldsymbol s}$. The ellipse has semiaxes of $S/2r$ and $S/2r\cos\vartheta$ and an area of $\mathrm d^2\hat{\boldsymbol s}=\pi S^2/4r^2\cos\vartheta$. Ocean waves, however, increase the size of this glint spot. As discussed by \citet{Williams2008}, the reflection signal does not significantly change as long as the ocean is wider than the spot. In the limit $S/r\longrightarrow 0$, the image of the star becomes a point (except for $\hat{\boldsymbol s}\bullet\hat{\boldsymbol r}=0$, which is the case of grazing reflection). The conclusion is that, in the regime $S\ll r\ll L$, the reflection signal equals the surface reflectivity at the point of reflection and thus probes the surface along a curve in the band
with polar values $\tfrac{\pi}{4}-\tfrac{1}{2}\theta\leq\mu\leq\tfrac{\pi}{4}+\tfrac{1}{2}\theta$. Close to edge-on observation $\theta\lesssim\frac{\pi}{2}$, the entire northern hemisphere is probed because, near inferior conjunction, the glint spot quickly moves from the $\mu=\frac{\pi}{2}$ to $\mu=0$. If we call ${\cal R}(\vartheta)$ the ocean reflectivity for incidence angle $\vartheta$,  the signal is
\[
f_\mathrm{L}(t) = {\cal R}(\vartheta(t))M_\mathrm{L}(\hat{\boldsymbol s}(t),t).
\]
Here and $M_\mathrm{L}(\hat{\boldsymbol s},t)$ is now the (rotating) ocean map. When taking a fixed vector $\hat{\boldsymbol s}$ on the surface, this map has the Fourier series $M_\mathrm{L}(\hat{\boldsymbol s},t) = \sum_{m=0}^\infty {\rm Re}(\mathrm e^{\mathrm im\Omega t}M^m(\hat{\boldsymbol s}))$. The glint point (in the nonrotating system) varies periodically over one year as $\hat{\boldsymbol s}(t+2\pi/\omega)=\hat{\boldsymbol s}(t)$. When we take this periodicity  into account, the coefficients for the rotation map are $M^m(\hat{\boldsymbol s}(t)) = \sum_{n=0}^\infty \mathrm e^{\mathrm in\omega t}M^m_n$, so that the net function $M_\mathrm{L}(\hat{\boldsymbol s}(t),t)$ is a quasiperiodic function of $t$. The reflectivity ${\cal R}$ is a function of time via $\vartheta(t)$, and also varies also annually. Therefore, the light curve for ocean glint, is a product of a periodic with a quasiperiodic function, which is quasiperiodic and has the form (\ref{quasiperiodic}), just like the light curve for a Lambertian map.

\begin{figure}
\centering
\resizebox{\hsize}{!}{
\includegraphics{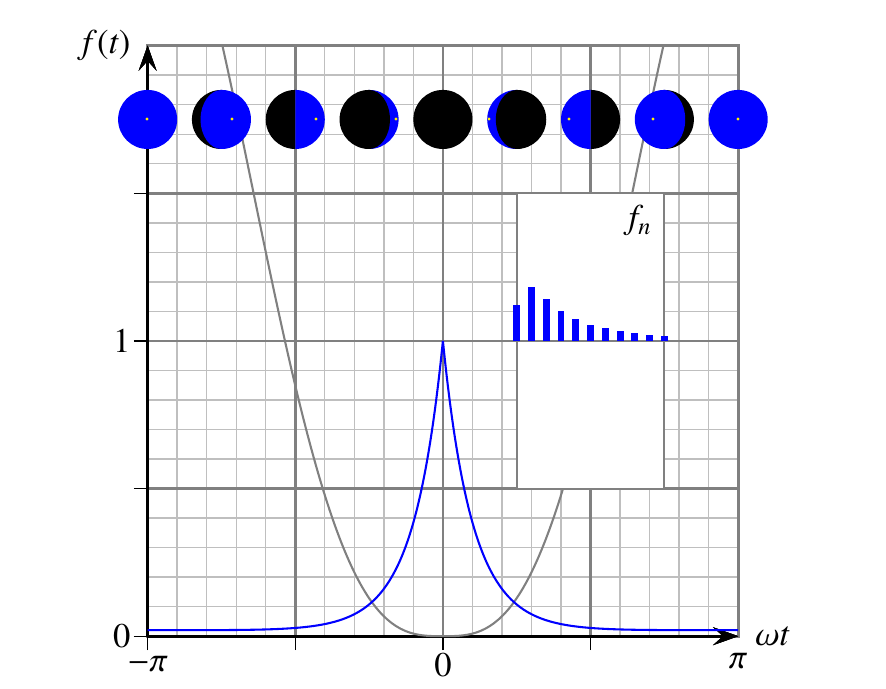}
}
\caption{\label{waterworld}
Standard Fresnel light curve (\ref{water}) from ocean glint for edge-on view (in blue; the Lambertian scattering signal is shown for comparison in gray.) The inset shows the spectrum. In the visuals, the glint is yellow.}
\end{figure}

\subsection{Fourier coefficients for ocean glint}
In this section, we derive a compact expression for the ocean glint, the ideal Fresnel curve, and calculate its Fourier peaks.
The reflection by a surface of water depends on the angle of incidence $\vartheta$. The reflection coefficients for $\mathrm s$-polarized light and $\mathrm p$-polarized light are given by the Fresnel equations \citep{Hecht2002}
\begin{align*}
{\cal R}_\mathrm{s}(\vartheta) &= \bigg(\frac{\sqrt{n^2-\sin^2\vartheta}-\cos\vartheta}{\sqrt{n^2-\sin^2\vartheta}+\cos\vartheta}\bigg)^2
, \\
{\cal R}_\mathrm{p}(\vartheta) &= \bigg(\frac{n^2\cos\vartheta-\sqrt{n^2-\sin^2\vartheta}}{n^2\cos\vartheta+\sqrt{n^2-\sin^2\vartheta}}\bigg)^2
.
\end{align*}
Taking $n=4/3$ for the index of reflection for water (or liquid methane), and assuming that the stellar light is unpolarized (the relative $\mathrm s$- and $\mathrm p$-intensities are $\frac{1}{2}$ and $\frac{1}{2}$) the net signal from a smooth water surface is
\begin{equation}
f = \frac{{\cal R}_\mathrm{s}+{\cal R}_\mathrm{p}}{2} = \frac{49(9-11\cos^2\vartheta+18\cos^4\vartheta)}{\Big(21+75\cos^2\vartheta+25\cos\vartheta\sqrt{7+9\cos^2\vartheta}\Big)^2} . 
\label{ftheta}
\end{equation}
For a planet in orbit, with position vector (\ref{r}), observed from direction (\ref{o}), the cosine of the incidence angle varies with time as
\[
\cos\vartheta(t) = -\hat{\boldsymbol r}\bullet\hat{\boldsymbol s} = \hat{\boldsymbol s}\bullet\hat{\boldsymbol o} =
\frac{\sqrt{1-\sin\theta\cos(\omega t-\phi)}}{\sqrt{2}}
.
\]
This must be substituted in (\ref{ftheta}) to get the light curve for a water world (with ocean map $M_\mathrm{L}(\hat{\boldsymbol s},t)=1$). The signal is significant if the observer is in the plane of the orbit, for $(\theta,\phi)=(\frac{\pi}{2},0)$, and $\cos\vartheta(t) = \sqrt{1-\cos\omega t}/\sqrt{2}= |\sin\tfrac{1}{2}\omega t\,|$. For this  case, we obtain the standard Fresnel light curve, which is
\begin{equation}
f_\mathrm{L}(t) = 
\frac{98(16-7\cos\omega t+9\cos^2\omega t)}{\displaystyle \Big(117-75\cos\omega t+25\sqrt{46-18\cos\omega t} |\sin\tfrac{1}{2}\omega t\,|\Big)^2}
.
\label{water}
\end{equation}
The glint spikes at $t=0$, for grazing light rays, when $\boldsymbol r$ and $\boldsymbol o$ are in the same direction; see Fig.\ \ref{waterworld}. The width extends to about a quarter of a year. The numerical values of the Fourier peaks are
\begin{align}
f_0 &= 0.12032 ,
\label{fwater} \\
f_1 &= 0.181 , \quad f_2 = 0.141 , \quad f_3 = 0.102 ,
\nonumber \\
f_4 &= 0.075 , \quad f_5 = 0.055 , \quad f_6 = 0.042 , \quad \mathrm{and}
\nonumber \\
f_n &\approx 50 \Big/ 3\sqrt{7}\pi n^2 \quad \mathrm{if}\quad n \longrightarrow \infty
.
\nonumber
\end{align}
Figure 4 shows the ratios of peaks that do not occur for a Lambertian planet. Observation of these values would strongly indicate that the planet is covered with oceans, composed of water or methane. The coefficients are positive and decay with an inverse square as $f_n\sim n^{-2}$, which is characteristic for functions that are differentiable except at one point (in this case at $t=0$). Many peaks can come out of the noise, in theory, for $|f_n|\gg 10^{-3}f_0$. However, the actual sharpness of the $t=0$ peak, will limit the validity of our calculated numerical values of $f_n$.

\subsection{Slushball Earth}
An interesting situation occurs for a planet that has an ocean circumventing the equator, and  is  covered with ice everywhere else. For Earth, this is called slushball Earth \citep{Hoffman1998,Fairchild2007,Macdonald2010};
it is the hypotheses that during the periods of global iceage, almost all of the oceans were covered with ice sheets, except for a narrow band at the equator. For an edge-on view, the signal has a maximal intensity: there is the scattering ice with nearly unit albedo and there is a maximal glint effect. The planet can at the same time be found in a transiting orbit, i.e.\ eclipsing the host star. The light curve is found with the summation of (\ref{snow}), (\ref{water}), (\ref{transit}). A Fourier coefficient is the sum of the expressions (\ref{fsnow}), (\ref{fwater}), (\ref{ftransit}) that follow.

\section{Polar caps}
\label{SecIV}
We describe the situation with polar caps as 
\begin{equation}
M_\mathrm{I}(\hat{\boldsymbol s}) = H(|\hat{\boldsymbol s}\bullet\hat{\boldsymbol n}|-\cos\alpha) , \quad
M_\mathrm{L}(\hat{\boldsymbol s}) = 1 - M_\mathrm{I}(\hat{\boldsymbol s}) ,
\label{H}
\end{equation}
using the Heaviside step function $H$, assuming that the caps are constant in size. Seasonal growth and decay of the caps is assumed to be small, which is reasonable for larger planets or small obliquity. Hence, $M_\mathrm{I}=0$ for $-\cos\alpha<\hat{\boldsymbol s}\bullet\hat{\boldsymbol n}<\cos\alpha$, and unity otherwise.  (An albedo factor and a cloud reduction factor could be included.) The border of the north polar icecap is found where
\begin{equation}
\cos\alpha = \hat{\boldsymbol s}\bullet\hat{\boldsymbol n} = \rho\sin\beta\cos\nu + \sqrt{1-\rho^2} \cos\beta
. \label{border}
\end{equation}
We assume $\beta\leq\alpha$, so that point $s\boldsymbol k$ lies on the north polar cap and we assume  $\alpha+\beta\leq\frac{\pi}{2}$, so that the tropical latitudes are ice free. The radial coordinate $\rho$ for the polar cap edge {\rm cap}, as a function of $\nu$, denoted $\rho_\mathrm{n}$, is then the positive solution of the quadratic equation (\ref{border}). One may write this equation and this solution as
\begin{align}
\rho^2 &= \frac{\cos^2\beta-\cos^2\alpha+2\rho\cos\alpha\sin\beta\cos\nu}{\cos^2\beta+\sin^2\beta\cos^2\nu}
, \label{rho}\\
\rho_\mathrm{n}(\nu) &= \frac{\cos\alpha\sin\beta\cos\nu + \cos\beta\sqrt{\sin^2\alpha-\sin^2\beta\sin^2\nu}}{\cos^2\beta+\sin^2\beta\cos^2\nu}
. \nonumber
\end{align}
The border of the south-polar cap is found at $\rho_\mathrm{s}(\nu)=\rho_\mathrm{n}(\nu+\pi)$.

\begin{figure}
\centering
\resizebox{\hsize}{!}{
\includegraphics{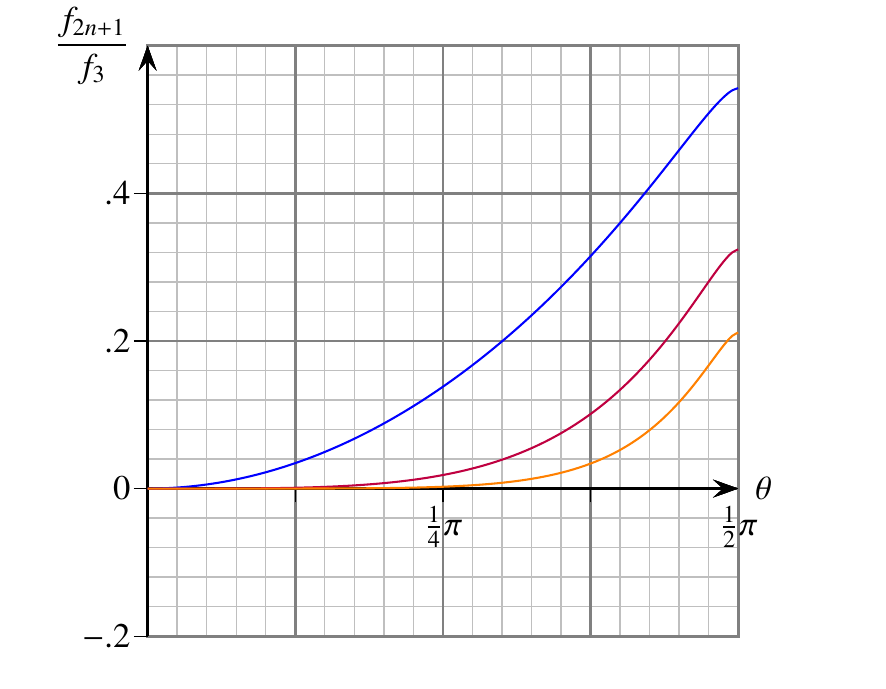}
}
\caption{\label{f5f3}
Harmonic ratios $f_5/f_3$ (blue), $f_7/f_3$ (purple), and $f_9/f_3$ (orange) for a planet with permanent ocean-glint. The glint spot of a water world with polar caps is continuously visible for $\theta>2\alpha+2\beta-\frac{\pi}{2}$.
}
\end{figure}

\begin{figure}
\centering
\resizebox{\hsize}{!}{
\includegraphics{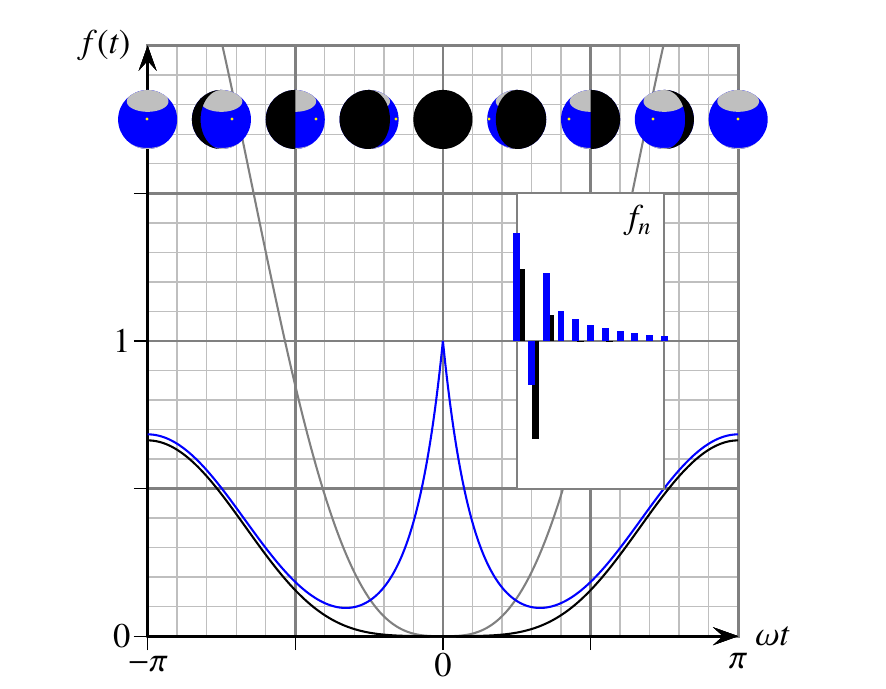}
}
\caption{\label{frontview}
Light curve of planet with polar caps for edge-on observation, $(\theta,\phi)=(\frac{\pi}{2},0)$, with and without glint (blue and black curve). The caps extend to $\alpha=\frac{\pi}{4}$ and the axial tilt is toward the observer with $\beta=\frac{\pi}{6}$. Inset shows the spectra, also with and without glint (blue and black).
}
\end{figure}

\subsection{Case I: edge-on view}
\label{SecIVA}
\begin{figure}
\centering
\resizebox{\hsize}{!}{
\includegraphics{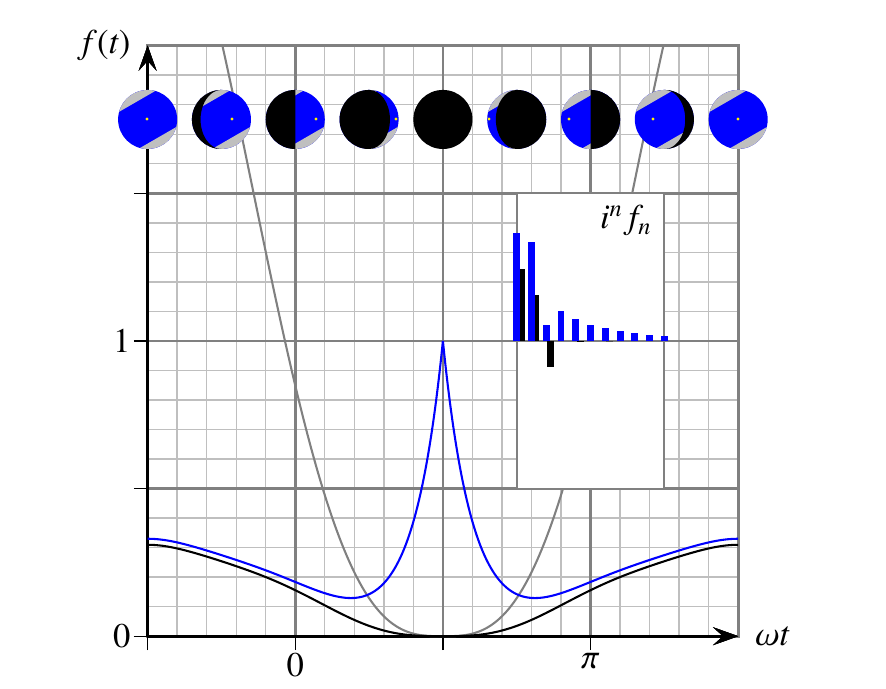}
}
\caption{\label{sideview}
Light curve of planet with polar caps  for edge-on observation, $(\theta,\phi)=(\frac{\pi}{2},\frac{\pi}{2})$, as in Fig.\ \ref{frontview}, except with the axis orthogonal to the observation direction.
}
\end{figure}

Consider the case that the orbital plane is observed edge on, when $\theta=\frac{\pi}{2}$. This important situation arises for extrasolar planets that were discovered with the transit method. During an eclipse, the planet passes between its parent star and Earth, so we observe the orbital plane edge on. The limb is no longer given by (\ref{limb}), but we find that  $\nu=\phi\pm\frac{\pi}{2}$. Between these values, for $\phi-\tfrac{\pi}{2} \leq \nu \leq \phi+\tfrac{\pi}{2}$, the pattern function is
\begin{equation}
\frac{g(\nu)}{2} = \bigg(\frac{4}{3} - \frac{2+\rho_\mathrm{n}^2}{3}\sqrt{1-\rho_\mathrm{n}^2} - \frac{2+\rho_\mathrm{s}^2}{3}\sqrt{1-\rho_\mathrm{s}^2}\bigg)\cos(\nu-\phi)
.
\label{gnuside}
\end{equation}
Substitution of (\ref{H}) in (\ref{g1}-\ref{g2n}) yields the Fourier coefficients
\begin{align*}
g_1 &= \frac{2}{\pi} \int\limits_{-\pi}^\pi \mathrm d\nu\, \mathrm e^{-\mathrm i\nu} \bigg( \frac{2}{3} - \frac{2+\rho^2_\mathrm{n}}{3}\sqrt{1-\rho^2_\mathrm{n}}\bigg) \cos(\nu-\phi)
, \\
g_{2n} &= \frac{2}{\pi} \int\limits_{-\pi}^\pi \mathrm d\nu\, \mathrm e^{-2\mathrm in\nu} \bigg( \frac{2}{3} - \frac{2+\rho^2_\mathrm{n}}{3}\sqrt{1-\rho^2_\mathrm{n}}\bigg) |\cos(\nu-\phi)|
.
\end{align*}
Here we need to substitute the cap-edge function (\ref{rho}). For small $\beta$, the coefficients are
\begin{align*}
g_1 &= 8\mathrm e^{-\mathrm i\phi}\frac{2+\cos\alpha}{3}\sin^4\frac{\alpha}{2} + \beta^2 \sin\alpha\sin 2\alpha \cos\phi
,  \\
g_2 &= 32\mathrm e^{-2i\phi}\frac{2+\cos\alpha}{9\pi}\sin^4\frac{\alpha}{2}  \\ 
&+ 2\beta^2\frac{15+5\mathrm e^{-2\mathrm i\phi}-\mathrm e^{-4i\phi}}{15\pi}\sin\alpha \sin 2\alpha
.
\end{align*}
If transits occur, the large signal (\ref{ftransit}) from the stellar light dip has to be carefully subtracted first. Figures \ref{frontview} and \ref{sideview} show light curves for an edge-on view in the respective cases  of $\phi=0$ and $\phi=\frac{\pi}{2}$.

\subsection{Case II: one polar cap in full view}
\label{SecIVB}
When the limb does not cross the icecaps, one cap is always completely visible and the other cap is entirely obscured. This allows us to go from $z$ to $\rho$ in equation (\ref{g}) and obtain, with substitution, (\ref{H}) that
\begin{align}
\frac{g(\nu)}{2} &= \int\limits_0^{\rho_\mathrm{n}}\mathrm d\rho\bigg(\frac{\rho^3\cos(\nu-\phi)}{\sqrt{1-\rho^2}}\sin\theta + \rho^2\cos\theta\bigg) ,
\label{fullview} \\
&= \bigg(\frac{2}{3} - \frac{2+\rho_\mathrm{n}^2}{3}\sqrt{1-\rho_\mathrm{n}^2}\bigg)\cos(\nu-\phi)\sin\theta + \frac{\rho_\mathrm{n}^3}{3}\cos\theta .
\nonumber
\end{align}
For small $\beta$ the first coefficients become
\begin{align*}
g_1 &= \mathrm e^{-\mathrm i\phi} \frac{16+8\cos\alpha}{3}\sin^4\frac{\alpha}{2}\sin\theta +
\beta\sin\alpha\sin2\alpha\cos\phi\cos\theta
, \\
g_2 &= \mathrm e^{-\mathrm i\phi}\beta\sin^3\alpha\sin\theta
.
\end{align*}

\subsection{Case III: zero obliquity}
\label{SecIVC}
\begin{figure}
\centering
\resizebox{\hsize}{!}{
\includegraphics{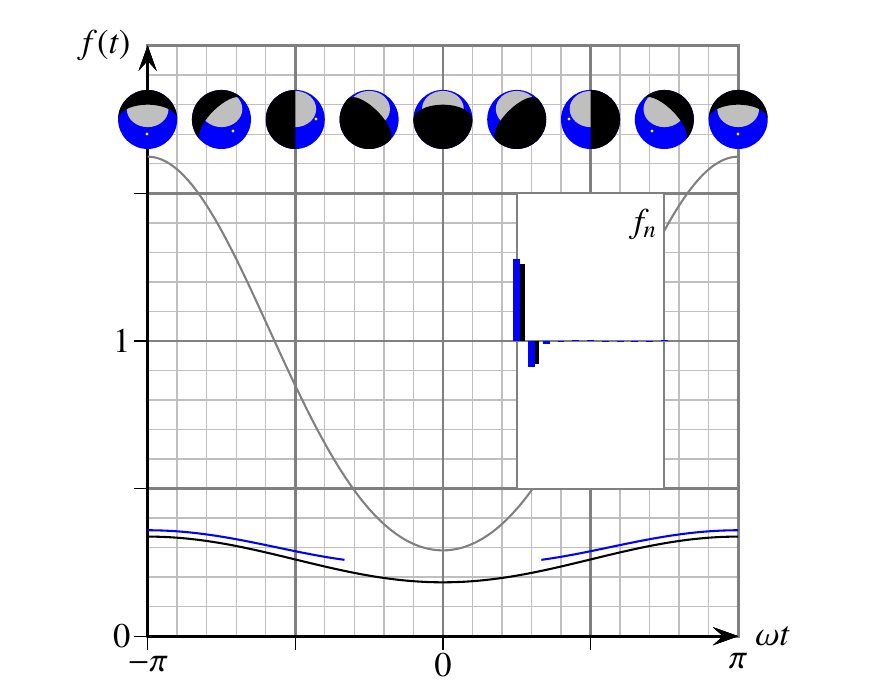}
}
\caption{\label{verticalaxis}
Light curve for planet with polar caps for inclined observation $\theta=\frac{\pi}{6}$. The caps extend to $\alpha=\frac{\pi}{4}$, and without axial tilt $\beta=0$. The north polar cap is completely in view, leading to a pure cosine. The glint is interrupted by the cap, but the effect is small. The gray curve is the Lambertian signal for this configuration.
}
\end{figure}

If the rotation axis is vertical, the case $\beta=0$, the symmetry about the $z$ axis is only broken in the observation direction (\ref{o}). Consider the case that the planet is observed from a direction that has full view onto the polar icecap. Then $\rho_\mathrm{n}=\rho_\mathrm{s}=\sin\alpha$, and (\ref{g},\ref{fullview}), with $\phi=0$, become
\[
\frac{g(\nu)}{2} = \bigg(\frac{2}{3}-\frac{2+\sin^2\alpha}{3}\cos\alpha\bigg)\cos\nu\sin\theta + \frac{\sin^3\alpha}{3}\cos\theta
.
\]
There are only two nonzero Fourier coefficients: $g_0$ and $g_1$. The relevant nonzero Fourier coefficient in the signal, $f_1=-g_1$ is reduced from $-4/3$ by the factor
\[
(4+2\cos\alpha)\sin^4\frac{\alpha}{2} \sin\theta
.
\]
Figure \ref{verticalaxis} shows an example of this case. A short calculation shows that the glint spot is on an open ocean for
\[
\cos\omega t < \frac{1}{\sin\theta}\Big( 1 - \frac{\cos^2\theta}{2\cos^2\alpha} \Big) .
\]

The case of zero obliquity with the observer inside the orbital plane gives the standard Lambertian curve (\ref{snow}) with harmonics (\ref{fsnow}), as in Fig.\ \ref{snowball}, except that Eqs.\ (\ref{snow}-\ref{fsnow}) must be multiplied with the cap size dependent reduction factor
\[
(4+2\cos\alpha)\sin^4\frac{\alpha}{2}
.
\]
Of course, for an edge-on observation of a water world with a polar cap, the ocean glint is maximal. This effect turns up in the odd harmonic amplitudes, $f_{2n+1}$, of the spectrum. When the reflection signal is subtracted from the spectrum, the Lambertian contribution remains, and the extension angle $\alpha$ of the polar caps can be determined from that contribution.

\subsection{Case IV: face-on view}
\begin{figure}
\centering
\resizebox{\hsize}{!}{
\includegraphics{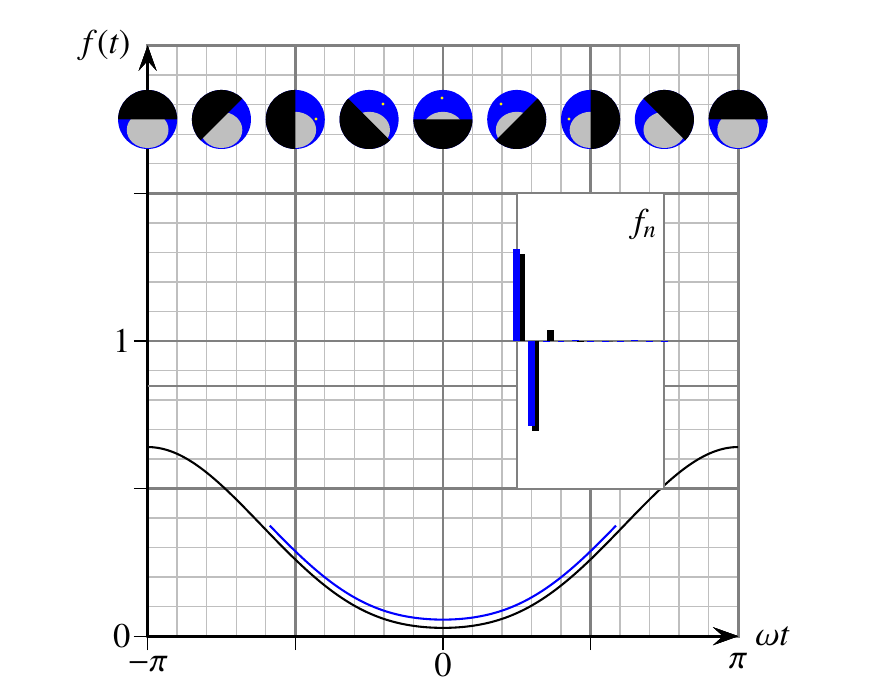}
}
\caption{\label{topview}
Polar-cap light curve (\ref{ftop}) for top view; inclination $\theta=0$. Cap extension is $\alpha=\frac{\pi}{4}$, obliquity is $\beta=\frac{\pi}{6}$. Graph and spectrum are clearly different from the situation in Fig.\ \ref{verticalaxis}. The glint is now constant and can be only be inferred from the caps interruptions. The gray line is the constant intensity from a homogeneous planet for top view.
}
\end{figure}

We consider, as a last concrete example, the situation that the observer has a top view onto the orbital plane, with  the entire northern polar cap also in view. For this case, $\theta=0$, and we require that $\alpha+\beta\leq\frac{\pi}{2}$. The pattern function (\ref{fullview}) for this situation in terms of the cap edge (\ref{rho}) is $g(\nu)=2\rho^3_\mathrm{n}/3$. To find the light curve itself, the Cartesian coordinates $(x,y)$ are most convenient as parameters for the northern hemisphere, where
\[
\hat{\boldsymbol s} = -\hat{\boldsymbol r}x + (\hat{\boldsymbol r}\times\boldsymbol k)y + \boldsymbol k \sqrt{1-x^2-y^2}
.
\]
The $x$ and $y$ coordinates are now the corotating coordinates, so that the day-night terminator circle found for $\hat{\boldsymbol s}\bullet\hat{\boldsymbol r}=0$ is at $x=0$. The surface element (\ref{element}) is
\[
\mathrm d^2\hat{\boldsymbol s} = \frac{1}{\sqrt{1-x^2-y^2}}\mathrm dx\mathrm dy,
\]
and the inner products are
${-\hat{\boldsymbol s}}\bullet\hat{\boldsymbol r} = x$,
$\hat{\boldsymbol s}\bullet\hat{\boldsymbol o} = \sqrt{1-x^2-y^2}$, which implies that $(-\hat{\boldsymbol s}\bullet\hat{\boldsymbol r})(\hat{\boldsymbol s}\bullet\hat{\boldsymbol o})\mathrm d^2\hat{\boldsymbol s}=x\mathrm dx\mathrm dy$. The border of the polar cap is located at $\hat{\boldsymbol s}\bullet\hat{\boldsymbol n}=$
\[
\cos\alpha = \sqrt{1-x^2-y^2}\cos\beta + (y\sin\omega t-x\cos\omega t)\sin\beta
.
\]
The quadratic form for the projection of the ellipse onto the $xy$ plane is
\[
(y^2+x^2-1)\cos^2\beta + (\cos\alpha-y\sin\beta\sin\omega t+x\sin\beta\cos\omega t)^2
= 0
.
\]
If we solve $y$ as a function of $x,$ we find solutions $y_+(x)$ and $y_-(x)$. The extreme values for $x$, where these two solutions connect, are
\[
x_\pm = \pm \sin\alpha\sqrt{1-\sin^2\beta\cos^2\omega t} - \cos\alpha\sin\beta\cos\omega t
.
\]
The surface integral in (\ref{formfactor}) can now be calculated using
\[
\int\limits_0^{x_+}\mathrm dx\, (y_+ -y_-)x = \frac{4\cos\beta\sin^2\alpha}{(x_+ -x_-)^2} \int\limits_0^{x_+}\mathrm dx\, \sqrt{x_+ - x} \sqrt{x-x_-} x
.
\]
The light curve is therefore given by
\begin{align}
f_\mathrm{I}(t) &=
\frac{12x_+^2-8x_+x_- +12x_-^2}{3\pi(x_+ -x_-)^2} \sqrt{|x_+x_-|} \cos\beta\sin^2\alpha
\label{ftop} \\
&+ \bigg( \frac{2}{\pi} \arctan\frac{x_+ +x_-}{2\sqrt{|x_+x_-|}} + 1 \bigg)(x_+ +x_-)\cos\beta\sin^2\alpha
.
\nonumber
\end{align}
The graph is plotted in Fig.\ \ref{topview} for polar caps that extend $45$ degrees and for obliquity of $30$ degrees .

\section{Tidally locked planet}
\label{SecV}
When a planet is tidally locked, the same side faces the host star at all times. The substellar point $\boldsymbol s=-\hat{\boldsymbol r}s$ is constantly directed toward the star and the point $\boldsymbol s=\hat{\boldsymbol r}s$ on the other side is always in the dark. Apart from moving cloud contributions, the surface pattern will be stationary in the corotating frame. A reflection signal from the planet is time dependent because an observer sees a rotating surface pattern with a constant illumination, instead of a time-varying illumination, as before. Since this is the special case $\Omega=\omega$, the general equations of Sec.\ \ref{SecIII} are valid, but now all coefficients $f_{n-m}^m$ contribute to the same $n\omega$-peak in (\ref{quasiperiodic}), for all $m$. We proceed, however, by defining a new pattern function. For this situation, we introduce rotating coordinates $(\nu,z)$ defined with
\[
\hat{\boldsymbol s} = ({-\hat{\boldsymbol r}}\cos\nu - \boldsymbol k\times\hat{\boldsymbol r}\sin\nu)\sqrt{1-z^2} + \boldsymbol k z
.
\]
In this coordinate frame, the surface maps $M_\mathrm{I}$ and $M_\mathrm{L}$ are only functions of $(\nu,z)$, and they are not functions of time. The planet will now have a rotationally symmetric map with respect to the time-varying $\hat{\boldsymbol r}$ axis, but not with respect to any fixed axis. One expects that for $\theta\neq 0$ and $\theta\neq\frac{\pi}{2}$, higher-order odd harmonics from both $M_\mathrm{I}$ and $M_\mathrm{L}$ are present in the spectrum. For edge-on observation $\hat{\boldsymbol o}=\boldsymbol i$, however, a similar symmetry as for a banded planet with a fixed rotation axis arises. We now have
\[
-\hat{\boldsymbol s}\bullet\hat{\boldsymbol r} = \sqrt{1-z^2}\cos\nu , \quad
\hat{\boldsymbol s}\bullet\hat{\boldsymbol o} = {-\sqrt{1-z^2}}\cos(\omega t+\nu)
.
\]
Equation (\ref{formfactor}) may therefore be expressed in terms of a pattern function $g$ for a locked planet, according to
\begin{align*}
f(t) &= \frac{4}{\pi} \!\int\limits_{\frac{\pi}{2}-\omega t}^{\frac{3\pi}{2}-\omega t} \mathrm d\nu\, |\cos(\omega t+\nu)| \frac{g(\nu)}{2}
,
\\
\frac{g(\nu)}{2} &= \cos\nu \int\limits_{-1}^1 \mathrm dz\, (1-z^2)M_\mathrm{I}(\hat{\boldsymbol s}) , \quad {\rm for} \quad -\frac{\pi}{2} \leq \nu \leq \frac{\pi}{2}
,
\end{align*}
and $g(\nu)$ is zero for the dark side of the planet, $|\nu|>\frac{\pi}{2}$. If we take the same definition (\ref{gn}), then the relation (\ref{fg}) for $m=0$, between the harmonics in the signal function $f$ and the pattern function $g$ is valid.

\begin{figure}
\centering
\resizebox{\hsize}{!}{
\includegraphics{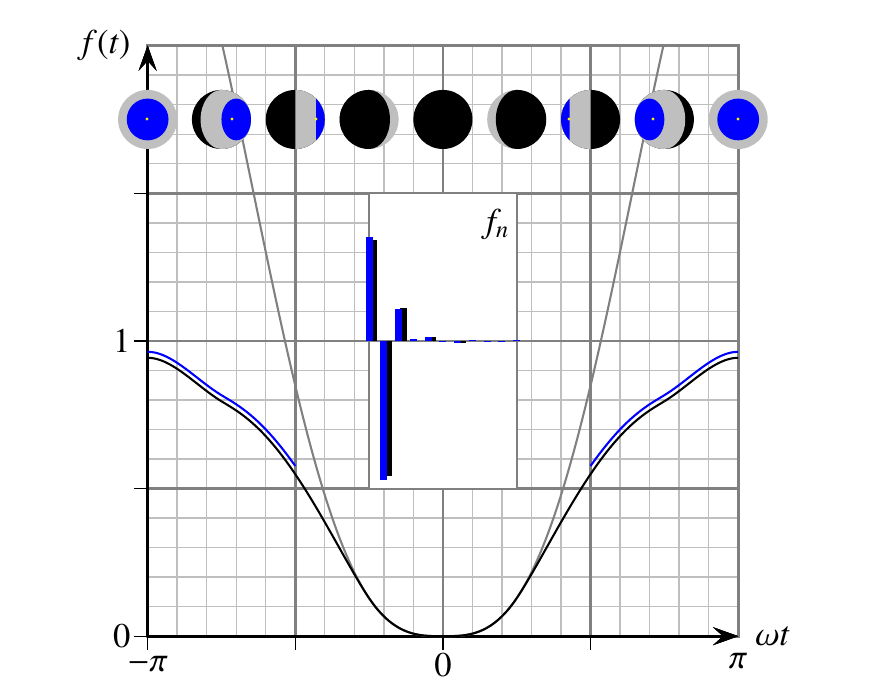}
}
\caption{\label{eyeball}
Light curve (\ref{eyeball1}-\ref{eyeball2}) for an eyeball planet with circular ocean extending over $\alpha=\pi/4$, with and without glint (blue and black), as given in Eq.\ (\ref{water}). The view is edge-on. The snowball light curve (gray) is for comparison.}
\end{figure}

\begin{figure}
\centering
\resizebox{\hsize}{!}{
\includegraphics{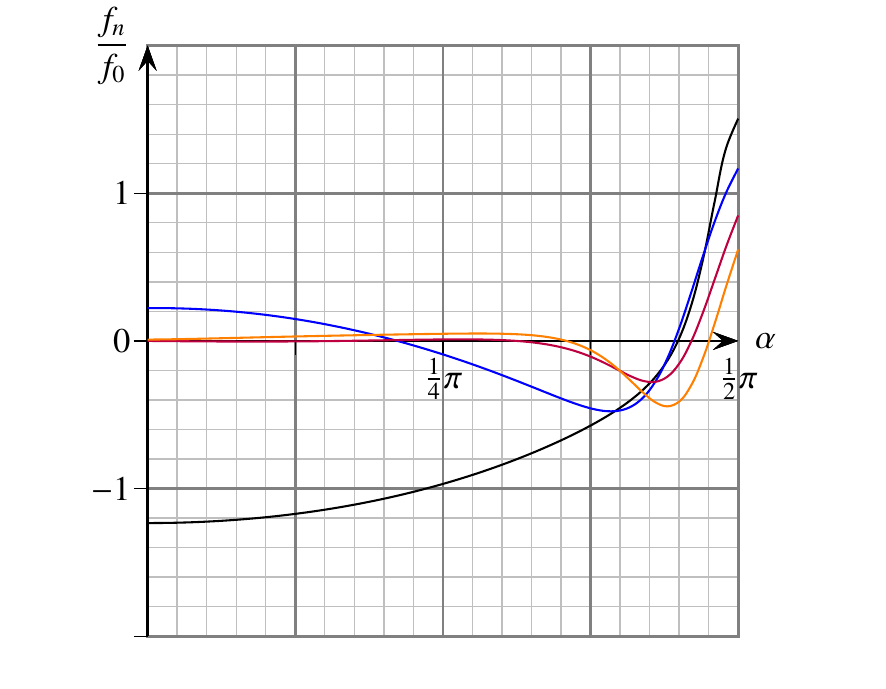}
}
\caption{\label{f1f2f3f4}
Peak ratios $f_1/f_0$ (black), $f_2/f_0$ (blue), $f_3/f_0$ (purple), and $f_4/f_0$ (orange) for an eyeball planet viewed edge-on plotted as a function of the ocean extension angle $\alpha$. Ocean glint is neglected.}
\end{figure}

\subsection{Eyeball worlds}
Consider now a tidally locked planet with open water. At the star-facing side,  a circular ocean exists with an extension angle $\alpha$ \citep{Pierrehumbert2011}. The rest of the planet is covered in ice, as depicted in the visuals of Fig.\ \ref{eyeball}. Because of oceanic heat transport, according to \citep{Hu2014}, the ocean shape may deviate from a circle for the relevant case of a locked planet near an M-type star. In our model, the edge is now located for $-\hat{\boldsymbol s}\bullet\hat{\boldsymbol r}=\cos\alpha$. We therefore assume the following form:
\[
M_\mathrm{I}(\hat{\boldsymbol s}) = H({-\hat{\boldsymbol s}}\bullet\hat{\boldsymbol r}-\cos\alpha) , \quad
M_\mathrm{L}(\hat{\boldsymbol s}) = 1 - M_\mathrm{I}(\hat{\boldsymbol s})
.
\]
We equate $g$ and $f$. We have for $\alpha\leq|\nu|\leq\frac{\pi}{2}$ that $g(\nu)=(8/3)\cos\nu$ and for $|\nu|<\alpha$ that $g(\nu)=$
\[
4\cos\nu \int\limits_0^{\frac{\cos\alpha}{\cos\nu}}\frac{\rho^3 \mathrm d\rho}{\sqrt{1-\rho^2}}
= \frac{8}{3}\cos\nu - \frac{4}{3} \Big( 2 + \frac{\cos^2\alpha}{\cos^2\nu}\Big) \sqrt{\cos^2\nu-\cos^2\alpha}
.
\]
At $\nu=\pm\alpha$ the pattern function has an infinite respective left- or right-derivative, leading to $g_n\sim n^{-3/2}$.

For $0<\omega t<\frac{\pi}{2}-\alpha$ the ocean is not visible, and $f$ equals the snowball signal (\ref{snow}). For $\frac{\pi}{2}-\alpha<\omega t<\frac{\pi}{2}+\alpha$ the ocean is partly visible, and the phase curve is
\begin{align}
f_\mathrm{I}(t) =& \frac{8\sin\omega t-8\omega t\cos\omega t}{3\pi}
\nonumber \\
& -\frac{8}{3\pi} \int\limits_{\frac{\pi}{2}-\omega t}^{\alpha}\mathrm d\nu\, |\cos(\omega t+\nu)| \Big( 2 + \frac{\cos^2\alpha}{\cos^2\nu}\Big) \sqrt{\cos^2\nu-\cos^2\alpha}
\nonumber \\
=& \frac{8\sin\omega t}{3\pi}
\label{eyeball1} \\
& - \frac{8}{3\pi}\Big( \omega t - \arccos\frac{\cos\omega t}{\sin\alpha} + \cos^3\alpha\arccos\frac{\cot\omega t}{\tan\alpha}\Big)\cos\omega t
\nonumber \\
& - \frac{8}{3\pi}\sqrt{\sin^2\omega t-\cos^2\alpha}\cos^2\omega t
- \frac{8}{3\pi}\Big(\sin^2\omega t-\cos^2\alpha\Big)^{3/2}
\nonumber
.
\end{align}
For $\omega t>\frac{\pi}{2}+\alpha$, the entire ocean is visible, and hence we need to subtract the above integral over the entire interval $[-\alpha,\alpha]$. We find
\begin{equation}
f_\mathrm{I}(t) = \frac{8\sin\omega t-8\omega t\cos\omega t}{3\pi} + (1-\cos^3\alpha)\frac{8\cos\omega t}{3}
\label{eyeball2}
.
\end{equation}
For $\omega t>\pi-2\alpha$, the glint spot is visible and a contribution $f_\mathrm{L}$, given by (\ref{water}),  must be added to the signal $f$. This time signal, shown Fig.\ \ref{eyeball}, has a peculiar wiggle in the interval when the ocean is in full view. Clearly, the pattern function $g$ is most useful for evaluation of the spectrum. The leading term in the large $n$ expansion is
\[
f_n \approx \sqrt{2\sin 2\alpha}(-1)^n\cos(2n\alpha-\tfrac{3\pi}{4}) \Big/ \pi^{3/2}n^{7/2} \quad {\rm if}\quad n\longrightarrow \infty
.
\]
If one wishes to determine $\alpha$ from the Fourier peaks, and verify that the signal is from an eyeball planet, the curves in Fig.\ \ref{f1f2f3f4} can be used.

\begin{figure}
\centering
\resizebox{\hsize}{!}{
\includegraphics{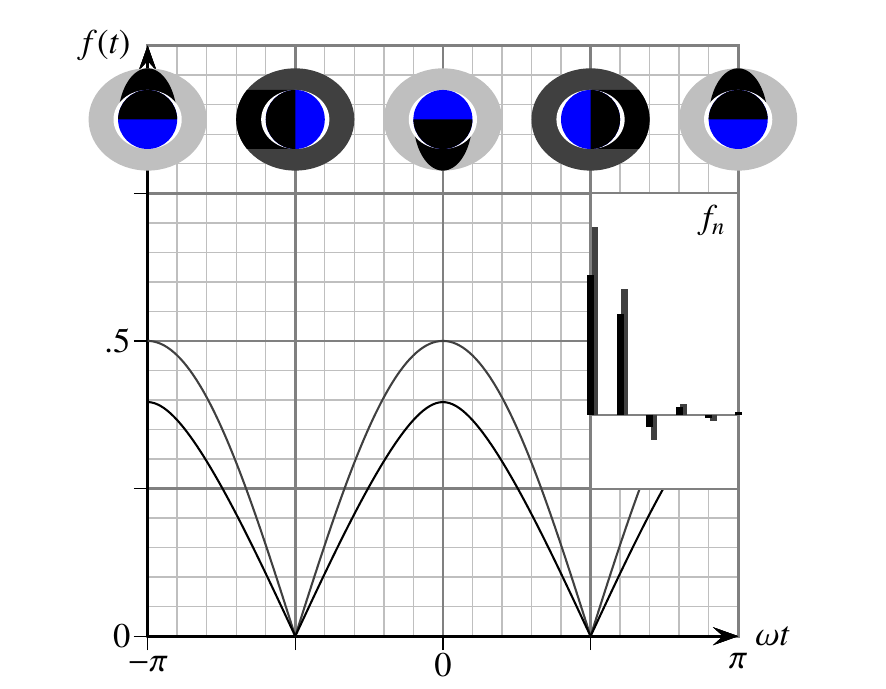}
}
\caption{\label{rings}
Light curves (\ref{ring}), (\ref{shadowring}) from Saturn-like rings with obliquity $\beta=\frac{\pi}{4}$ and ${\cal R}={\cal T}=\frac{1}{2}$ (so that rings can be observed lit-from-behind) and zero inclination; with and without planet-casting shadows (black and gray). Then, shown in the inset $f_1=0$.  For orbital phases $\omega t=\pm\frac{\pi}{2}$, the rings are edge-on illuminated and the signal is zero.}
\end{figure}

\section{Saturn-like rings}
\label{SecVI}
Consider a Saturn-like planet, a planet with a large disk-shaped ring system. We model the ring system as a homogeneous and the thin disk in the plane with normal vector $\hat{\boldsymbol n}$ as in (\ref{n}). With the inner radius $b$ and the outer radius $c$, the total surface area is $\pi c^2-\pi b^2$. We study configurations with $s\leq (\hat{\boldsymbol n}\bullet\hat{\boldsymbol o})b$, so that the planet does not block a part of the observer's view on the ring. Light scatters off the ring with reflectivity ${\cal R}$. For a thin disk, light can pass through so that the rings are also visibly back lit \citep[see][]{Porko2005}. The light that comes out the other side is again scattered diffusive, with transmitivity ${\cal T}$. This approach requires the disk to be optically thick \citep[for optically thin rings, see][]{Arnold2004}. We factorize the net intensity $F(t)$ for the light scattered off the ring according to
\[
F(t) = \frac{\pi c^2-\pi b^2}{4\pi r^2} ({\cal R}+{\cal T}) I  f(t) (\hat{\boldsymbol n}\bullet\hat{\boldsymbol o})\sin\beta,
\]
with $f(t)$ independent of the ring absorption $1-{\cal R}-{\cal T}$, and also with the dependence on observation direction $\hat{\boldsymbol o}$ factored out, as well as the sine of the ring tilt angle $\beta$.

Consider first the simplified case where the planet is much smaller than the ring disk, hence $s\ll c$, so that the planet's shadow is unimportant. For the interval $|\omega t|<\frac{\pi}{2}$ where the illuminated face has normal $\hat{\boldsymbol s}=-\hat{\boldsymbol n}$ directed away from the observer and only transmitted light is visible with intensity ${\cal T}$, whereas for the other half orbit, the illuminated face has $\hat{\boldsymbol s}=\hat{\boldsymbol n}$, with the scattered intensity ${\cal R}$. The phase curve for the rings is then
\begin{align}
f_\mathrm{I}(t) &= \frac{|\cos\omega t\,|+\cos\omega t}{2} \frac{{\cal T}}{{\cal R}+{\cal T}} + \frac{|\cos\omega t\,|-\cos\omega t}{2} \frac{{\cal R}}{{\cal R}+{\cal T}}
\nonumber \\
&= \frac{{\cal T}-{\cal R}}{2{\cal R}+2{\cal T}}\cos\omega t + \frac{1}{2}|\cos\omega t\,|
.
\label{ring}
\end{align}
The Fourier coefficients are
\begin{equation}
f_0 = \frac{1}{\pi} , \quad 
f_1 = \frac{{\cal T}-{\cal R}}{2{\cal R}+2{\cal T}} , \quad
f_{2n} = \frac{-(-1)^n}{4n^2-1}\, \frac{2}{\pi} , \quad f_{2n+1} = 0
.
\label{fring}
\end{equation}
The case of opaque rings, for ${\cal T}=0$, actually produces the same signal as polar caps in the small-cap size limit $\alpha\longrightarrow 0$. When $0<\alpha\ll\beta$ the polar cap is located inside the polar circle of latitude. The surface integral (\ref{formfactor}) is reduced to a very small circle around $\hat{\boldsymbol n}$ so that the factor $(\hat{\boldsymbol s}\bullet\hat{\boldsymbol o})$ can be considered constant. The signal is a cosine function when this cap experiences polar day, and the signal is zero during polar night. This situation is opposite to the cases from Sect.\ \ref{SecIII} where we only considered $\alpha>\beta$.

\begin{figure}
\centering
\resizebox{\hsize}{!}{
\includegraphics{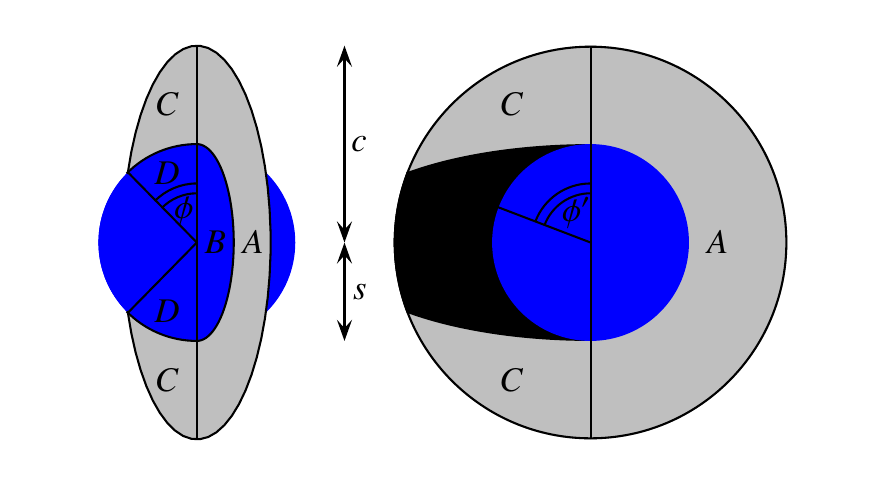}
}
\caption{\label{saturn}
Ringed planet, with planet radius $s$ and outer ring radius $c$ viewed from the direction $-\hat{\boldsymbol r}$ of the star, defining $\phi$ (left side) and  from the polar north $\hat{\boldsymbol n}$, defining the angle $\phi'$ (right side). The illuminated area consists of $A$ plus $C$. On the left, $A$ is the difference between two half ellipses with areas $A+B$ and $B$; and $C$ the projected circle segment $C+D$ minus a circle segment $D$ (with respective areas $\phi' c^2\sin\beta$ and $\phi s^2$).}
\end{figure}

\subsection{Shadow cylinder}
We discuss, as a final application, the situation where the planet also casts a shadow on the rings. In this case the planet itself  and  the shadow cast by the rings onto the planet are neglected, compare \citet{Arnold2004}. This shadow is obviously important when the planet is not small compared to the rings. Consider the case when the planet is so large that the shadow cylinder of the planet always intersects the ring and at any time part of the ring is dark. This means that for our calculation the obliquity is bounded by
\[
c\sin\beta \leq s \leq b\cos(\theta-\beta)
.
\]
We calculate the illuminated effective area using the plane orthogonal to $\boldsymbol r$. If we view the area from the position of the central star, this area is precisely the region not obscured by the planet disk. This geometry appears in Fig.\ \ref{saturn} (on the left). We first imagine that the ring starts at the planet surface, for radii between $s$ and $c$. Since the real system has radii between $b$ and $c$, to obtain the required signal we must subtract  the same expression with $c$ replaced by $b $ from this initial result. Thus consider now first a ring running from $s$ to $c$. The cross section of the half of this ring  that is on the dayside of the planet is $(c^2-s^2)|\hat{\boldsymbol r}\bullet\hat{\boldsymbol n}|\pi/2$, since the dayside is fully illuminated. The illuminated part of the nightside half of the ring consists geometrically of two (identical) segments of the circle of radius $c$ with angle $\phi'$ projected into the outer ellipse, minus two circle segments of radius $s$ with angle $\phi$ (see Fig.\ \ref{saturn}). Hence, $\phi$ is the angle between the point where the shadow circle of the planet crosses the outer (projected) ellipse and the direction normal to $\hat{\boldsymbol n}$, whereas $\phi'$ is the angle between these points for a face-on view. This yields a cross-section $c^2|\hat{\boldsymbol r}\bullet\hat{\boldsymbol n}|\phi'-s^2\phi$. Substitution of the point on the circle in the equation for the outer ellipse gives
\[
\Big(\frac{s\cos\phi}{c}\Big)^2 + \Big(\frac{s\sin\phi}{c\hat{\boldsymbol r}\bullet\hat{\boldsymbol n}}\Big)^2 = 1
.
\]
It follows that the angle is given by
\[
\phi(c,t) = \arctan\sqrt{\displaystyle\frac{c^2-s^2}{s^2/\cos^2\omega t\sin^2\beta-c^2}}
.
\]
The relation between $\phi'$ and $\phi$ is $|\hat{\boldsymbol r}\bullet\hat{\boldsymbol n}|\tan\phi' = \tan\phi$. Hence, the second angle is given by
\[
\phi'(c,t) = \arctan\sqrt{\displaystyle\frac{c^2-s^2}{s^2-c^2\cos^2\omega t\sin^2\beta}}
.
\]
If we now add the contributions of the two ring halves, we obtain
\begin{align*}
f_\mathrm{I}(t) =  \frac{1}{2}|\cos\omega t| &+
\frac{1}{\pi} \frac{c^2\phi'(c,t)-b^2\phi'(b,t)}{c^2-b^2} |\cos\omega t|
\\
&- \frac{s^2}{c^2-b^2} \frac{\phi(c,t)-\phi(b,t)}{\pi\sin\beta}
,
\end{align*}
where we must remark that these equations are only valid for $|\omega t|\geq\frac{\pi}{2}$ and ${\cal T}=0$.
This is the situation when the rings are opaque
and only visible half the time. The general case is just a linear combination similar to (\ref{ring}). To lowest order in the axial tilt $\beta$, and for $b=s$, the result is simply proportional to Equation (\ref{ring}), but with an extra geometric factor
\[
\frac{1}{2} + \frac{c^2}{\pi(c^2-s^2)}\arccos\frac{s}{c} - \frac{s}{\pi\sqrt{c^2-s^2}}
.
\]
When (\ref{fring}) is multiplied with this same factor, the asymptotic behavior of $f_n$ for large $n$ is accurately described for arbitrary $\beta$. For the case of maximal visibility, and arbitrary $\beta$, which is $\hat{\boldsymbol n}=\hat{\boldsymbol o}$, $b=s$, and $c=s/\sin\beta$, the signal from the rings simplifies to $f_\mathrm{I}(t) =$
\begin{equation}
\bigg( \frac{1}{2} + \frac{1}{\pi\cos^2\beta}\arctan\frac{\cot\beta}{|\sin\omega t|}\bigg)|\cos\omega t| - \frac{\sin\beta}{\pi\cos^2\beta} \arctan\frac{\cos\beta}{|\tan\omega t|}
.
\label{shadowring}
\end{equation}
The graph appears in Fig.\ \ref{rings}. The signal is significantly weaker because for $b=s$, the planet shadow is large.

The spectrum of the signal from a planet with rings is of the form (\ref{cosine}) and is again characterized by the fact that higher order odd coefficients are all zero. For ${\cal T}={\cal R}$, also $f_1=0$. The dominant Fourier coefficient of (\ref{shadowring}) is a relatively short formula,
\[
f_2 = \frac{1}{6\pi^2\cos^2\beta} \bigg( 3\pi + \pi\cos 2\beta - \frac{4}{\tan\beta} - \frac{8\log\cos\beta}{\tan^3\beta} - 4\beta \bigg)
.
\]
The Fourier peaks  for rings behave asymptotically as $(-1)^{n+1}n^{-2}$ like (\ref{fring}), whereas for the planets the tail behaves as $n^{-4}$ like (\ref{fsnow}). This behavior allows one to discriminate between the signal from a thin disk and  a sphere.

\section{Conclusions}
When a planetary system is observed continuously over a number of years, the Fourier transform of the light signal provides a natural method of analysis. The immediate advantages are:
\begin{inparaenum}[(i)]
\item The glare from the host star (in observations for large planets near M-dwarfs without an occultor), or some residual glare, can be filtered out. Stellar noise at the orbital frequencies swamps the Fourier peaks for small $s^2/r^2$.
\item Atmospheric cloud cover that partially obscures the surface, is also partially filtered out, if clouds mainly contribute an a-periodic component in the signal.
\item Well-defined orbital and diurnal frequencies \citep[see][]{Palle2008} can be obtained from the peak-center positions.
\item Contributions of different planets can be identified, provided they are not in orbital resonance.
\item The asymptotic behavior of the peak-amplitudes, phase factors, and even-odd structure allows us to separate out and identify different contributions (reflective oceans, land or ice sheets, or rings) in the light signal. \item For observation time $T$ of $N$ orbits, the signal-to-noise ratio in the peaks is enhanced by the root of $N$, essentially because the Fourier transform is an averaging method.
\end{inparaenum}

How can one distinguish between planets with or without rings and identify the different configurations we consider here? Since none of the specific parameters are known in advance, one has to rely on general properties of a measured Fourier spectrum. To apply our results, an astronomer should first identify the spectral peaks from the intensity spectrum of a star system that belong to a single planet. If sidebands to the annual $n\omega$-peak pattern (a similar pattern shifted by $\Omega$ and $2\Omega$, etc.) are not found, either the planet is banded and has no longitudinal pattern, or the planet is tidally locked (so that $\Omega=\omega$). Then the astronomer needs to fix the orbital phase with respect to the inferior conjunction. For a transiting planet, this is at moment of  transit (peaking).

If the coefficients are all positive and decay as $f_n\sim n^{-2}$, this may indicate ocean glint, see Eq.\ (\ref{fwater}). If there are no higher-order odd peaks, $f_{2n+1}=0$, then there is no glint, which implies that the point of specular reflection is on land or behind a thick atmosphere, see Eq.\ (\ref{fg}). When the glint spot abruptly moves from ocean to a polar cap or back, this generally occurs when the glint effect is weak. We therefore expect that the $n^{-1}$ behavior due to such a discontinuous jump will only dominate in the far-end regions of the tail that cannot be resolved. If the spectrum (after subtracting any observed glint component) has an alternating even tail behaving as $f_{2n}\sim(-1)^{n+1}n^{-2}$, the signal is that of a scattering disk, either from a flat polar-cap (inside the polar circle) or from a ringed system. The value of $f_1$ can determine if the signal is light scattered by semitransparent rings, see Eq.\ (\ref{ring}). The peaks of an locked planet with a circular ocean decay in strength as $|f_n|\sim n^{-7/2}$. A generic Lambertian signal from a planet decays fast as $f_{2n}\sim n^{-4}$. If we assume a stellar-noise level of $10^{-3}$, the number of peaks that is expected to be resolved  is about $100$, $30$, $10$, and $6$, for a perfect glint spike, rings, and Lambertian scattering ice caps, respectively. If the photometry is not good enough to detect the fraction $10^{-3}$ of the average planet luminosity, of the order $s^2I/4r^2$, then fewer peaks can be found.

For observation that is not edge on the orbital plane, the spectrum obtains an extra factor of $\sin^{2n}\theta$. This exponential decay will limit the number of peaks that can be resolved, which will hamper identification. The effect of orbit eccentricity, see Eq.\ (\ref{keplercoef}) is also exponentially weak because $f_n\sim (\epsilon\mathrm e/2)^n$. This effect dominates only for zero inclination, and in case obliquity is zero or the planet is homogeneous (or at least the exposed region is). Any other periodic signal is caused by changing phases or rotating patterns on the planet surface. The expressions at the end of Sects.\ \ref{SecIVA}-\ref{SecIVB} for $g_1$, $g_2$ are simple enough to see whether the four parameters, $\theta,\phi,\alpha,\beta$, for a water world with polar caps fit the observation of the first two Fourier peaks.

If one isolated peak appears in the spectrum, this could indicate a homogeneous but large planet at such a small inclination that all other peaks are suppressed. However, if all the reflected light comes from a circle around the (north or south) pole that happens to be in full view, a pure cosine light curve is precisely what will be observed for a planet with zero axial tilt, see Sect.\ \ref{SecIVC}. A bright circularly symmetric albedo map around a planet's pole on a further absorbing surface, could indicate an exposed circular polar ice cap surrounded by oceans.

The main disadvantage of the Fourier transform is the need for many periods, hence a long continuous observation (in the optical range), but we think that it would still be worthwhile to single out a promising exoplanetary system as a candidate for study of the intensity spectrum.

\bibliographystyle{aa}
\bibliography{24992}{}

\newpage
\appendix
\section{Transits and observation gaps}
\label{AppA}
When planet transits and reflected light curves are observed in one signal, large amplitudes from the periodic transits are added on top of the weak reflection peaks. For Jupiters near dwarf stars, it may still be possible to separate the two components in the signal.

Observing from direction $\hat{\boldsymbol o}$, one sees a planet passing in front of the star disk for $|\boldsymbol r\times\hat{\boldsymbol o}|<S$. The transits are thus found (taking $\phi=0$ for convenience, and $r\ll S$, and $\theta\approx\frac{\pi}{2}$) in $t$-intervals for which
\[
r\sin|\omega t| < r\sin\omega\tau = \sqrt{S^2-r^2\cos^2\theta}
.
\]
Then the measured signal dips by $-s^2I_0/S^2$ (which is $-10^{-4}I_0$ for a twin earth). We are interested in the reflected signal from a planet. This is a factor $S^2/4r^2$ weaker ($10^{-6}$) than a transit signal and $s^2/4r^2$ weaker ($10^{-10}$) than the star. For a Jupiter close to a dwarf star, these ratio's are much more practical than for an earth-twin, and it may not be necessary to block the star.

Although the transit signal is of short duration (minutes), the combined signal can be detected continuously (for years). Consider the Fourier series of a rapid transit:
\begin{equation}
I(t) = I_0 - \frac{s^2I_0}{rS\pi}\sqrt{\displaystyle 1-\frac{r^2\cos^2\theta}{S^2}}\bigg( 1+ 2\sum_{n=1}^\infty \frac{\sin n\omega\tau}{n\omega\tau}\cos n\omega t \bigg)
.
\label{transit}
\end{equation}
The short duration of the eclipse gives rise to peak amplitudes of a fraction $s^2/rS$ of $I_0$ (of the order $10^{-7}I_0$). The Fourier coefficients of the transit signal are
\begin{equation}
f_n = -\frac{8r}{\pi S} \frac{\sin n\omega\tau}{n\omega\tau} \sqrt{\displaystyle 1-\frac{r^2\cos^2\theta}{S^2}}
\label{ftransit}
,\end{equation}
which are of the order $r/S$ compared to the coefficients in the phase signal, which are of order $1$. In other words, in the Fourier spectrum, the reflected signal is about a factor $S/r$ weaker ($10^{-3}$) than the transit signal. To observe the phase part in a combined phase-plus-transit signal, the noise from the transit must be less than the phase peak amplitudes. Our Sun has noise levels of $10^{-5}$ at (transit-frequencies of) $10^{-5}\mathrm{Hz}$, and a noise level $10^{-3}$ for (orbital frequencies) $10^{-6}\mathrm{Hz}$, see \citet{Aigrain2004}, so the large transit component does indeed pose a problem since the absolute noise will be large. Of course, when a planet is periodically blocking its star, the star is also blocking the planet every other half period. The corresponding dip in the signal is again a factor $S/r$ smaller.

The effect of discontinuous but periodic observation gaps on the Fourier peaks is different from that of a transit. For observation gaps, the continuous signal is multiplied with a periodic function, which is zero for the intervals $\sin|\omega t|<\sin\omega\tau $, whereas for a transit, such a function is added instead. Periodic interruptions at the inferior and major conjunction are inevitable for an edge-on observation, when the planet passes behind an occulter or within the telescope's inner working angle. The blocking durations $2\tau$ must obviously be short (compared to the period). In the short-gaps limit, the multiplicative function becomes $1-2\omega\tau \sum \mathrm e^{2\mathrm in\omega t}/\pi$. The even and odd peaks, $f_{2n}$ and $f_{2n+1}$ are shifted by $(\mp f(0)-f(\pi/\omega))\omega\tau/\pi$, in the approximation $n\omega\tau\ll 1$.

\begin{figure}
\centering
\resizebox{\hsize}{!}{
\includegraphics{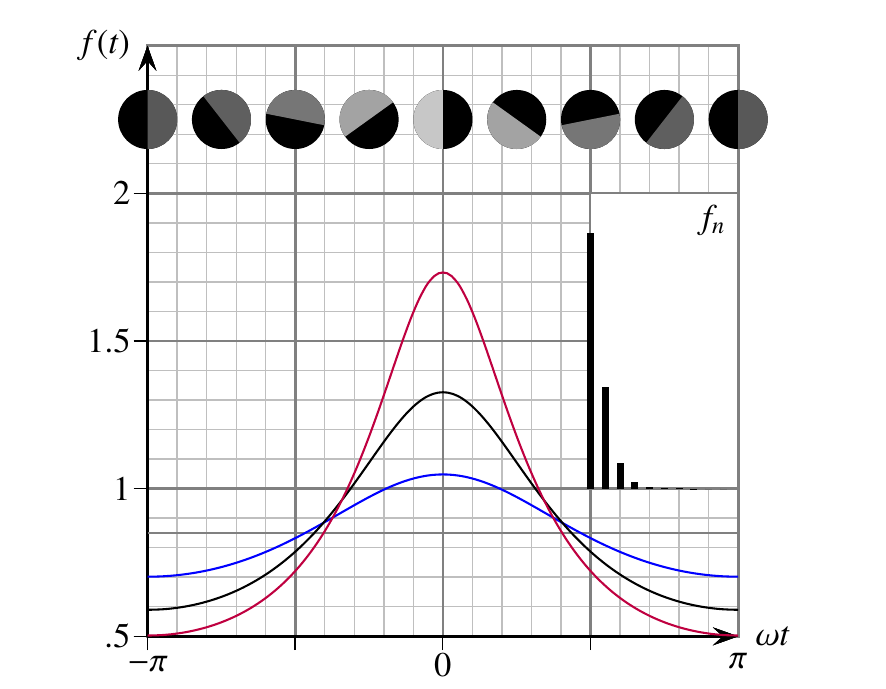}
}
\caption{\label{kepler}
Phase-light curves for top view on an eccentric planetary orbit. Values for the eccentricity are $\epsilon=0$ (gray), $\epsilon=.1$ (blue), $.2$ (black), and $.3$ (purple, highest peak intensity). The visuals and the inset are for $\epsilon=.2$.}
\end{figure}

\section{Kepler orbits}
\label{AppB}
For planets with noncircular, but eccentric orbits, the distance to the host star is a time-varying periodic function. As a result, the light intensity received by the planet is periodically modulated. We will calculate here the effect of (small) orbital eccentricity on phase light curves. The position vector is no longer given by the circle (\ref{r}) but with an ellipse with a semimajor axis $a$ along the $x$-direction and eccentricity $\epsilon$, given by
\[
\boldsymbol r(t) = x(t)\boldsymbol i + y(t)\boldsymbol j = (\boldsymbol i\cos\nu(t) + \boldsymbol j\sin\nu(t))r(t)
.
\]
The Cartesian coordinates $(x,y,z)=(x,y,0)$ are
\[
x(t) = a\cos\lambda(t) - a\epsilon , \quad y(t) = \sqrt{1-\epsilon^2}a\sin\lambda(t)
,
\]
and the polar coordinates $(r,\mu,\nu)=(r,\frac{\pi}{2},\nu)$ are given by
\[
r(t) = a - a\epsilon\cos\lambda(t) , \quad \cos\nu(t) = \frac{\cos\lambda(t)-\epsilon}{1-\epsilon\cos\lambda(t)}
.
\]
These orbital coordinates are expressed in the eccentric anomaly $\lambda$ \citep[usually denoted $E$, see][]{Goldstein1964}. This parameter $\lambda$ is (implicitly) defined as a function of  the orbital phase $\omega t$, from
\[
\omega t = \lambda - \epsilon\sin\lambda
.
\]
An \emph{expansion\/} in $\epsilon$ for the coordinates in the time variable can be obtained with
\[
\lambda(t) = \omega t + \epsilon\sin\Big( \omega t + \epsilon\sin\big( \omega t + \cdots \big) \Big)
.
\]

In the general case, the eccentricity of the orbit leads to a modulation factor $a^2/r^2(t)$ in the light curve. The simplest case arises for a homogeneous planet with inclination $\theta=0$. (Taking $M_\mathrm{I}=1$, $\epsilon=0$, the half-illuminated disk gives $f=\frac{8}{3\pi}$, neglecting the possible $45$ degrees glint effect.) The signal is
\[
f(t) = \sum_{n=0}^\infty f_n\cos n\omega t = \frac{8}{3\pi} \frac{a^2}{r^2(t)} = \frac{8}{3\pi(1-\epsilon\cos\lambda(t))^2}
.
\]
The graph appears in Fig.\ \ref{kepler}. Note that Mercury has $\epsilon=.2$.

We calculate the Fourier coefficients $f_n$ by first considering the function $1/r(t)$.
We use $\lambda$ as integration variable and require for this substitution
\[
\mathrm d(\omega t) = \mathrm d(\lambda - \epsilon\sin\lambda) = (1-\epsilon\cos\lambda)\mathrm d\lambda
,
\]
so that
\[
\frac{1}{r(t)}\mathrm d\omega t = \frac{1}{a}\mathrm d\lambda
.
\]
When the complex harmonic $\mathrm e^{\mathrm in\omega t}$ is expressed as a periodic function of $\lambda$, one obtains the generating function for the Bessel functions (of the first kind)
\[
\mathrm e^{\mathrm in\omega t} = \mathrm e^{\mathrm in\lambda}\mathrm e^{-\mathrm in\epsilon\sin\lambda} = \sum_{k=-\infty}^\infty \mathrm e^{\mathrm i(n-k)\lambda} J_k(n\epsilon)
.
\]
Combining the last two identities and integrating over a full period gives us the Fourier coefficients for $1/r(t)$, which are
\[
\frac{1}{2\pi}\int\limits_{-\pi}^{\pi}\mathrm e^{\mathrm in\omega t}\frac{1}{r(t)}\mathrm d\omega t
= \frac{1}{a}J_n(n\epsilon)
.
\]
Hence, the reciprocal distance has Fourier series
\[
\frac{1}{r(t)} = \frac{1}{a}\sum_{n=-\infty}^\infty J_n(n\epsilon)\mathrm e^{\mathrm in\omega t} = \frac{1}{a} + \frac{2}{a} \sum_{n=1}^\infty J_n(n\epsilon)\cos n\omega t
.
\]
We used $J_{-n}(-n\epsilon)=J_n(n\epsilon)$ to get the last identity. It is interesting that the Fourier coefficients of $r(t)$ and of $\cos\theta(t)$ are Bessel functions, although the functions $r(t)$ and $\cos\theta(t)$ themselves have no explicit expression in $t$. The Fourier coefficients of the all powers $r^N(t)$, with $N\geq-1$ of the radial coordinate $r(t)$ are finite combinations of Bessel functions too.

The luminosity of a planet is, however, inversely proportional to the square of the distance $r(t)$. We may use the convolution the series for $1/r(t)$ with itself to obtain
\[
\frac{1}{r^2(t)} = \frac{1}{a^2} \sum_{n=-\infty}^\infty \mathrm e^{\mathrm in\omega t} \sum_{k=-\infty}^\infty J_k(k\epsilon)J_{n-k}(n\epsilon-k\epsilon)
.
\]
Alternatively we may write
\[
\mathrm e^{\mathrm in\omega t}\frac{1}{r^2(t)}\mathrm d\omega t =\frac{1}{a^2} \mathrm e^{\mathrm in\lambda}\mathrm e^{-\mathrm in\epsilon\sin\lambda} \frac{1}{1-\epsilon\cos\lambda}\mathrm d\lambda
\]
and use the series
\[
\frac{1}{1-\epsilon\cos\lambda} 
= \frac{1}{\sqrt{1-\epsilon^2}} \sum_{k=-\infty}^\infty \mathrm e^{ik\lambda} \varepsilon^{|k|}
.
\]
The parameter $\varepsilon$ is a new measure of the orbital eccentricity (also in the interval $[0,1]$) defined with $\varepsilon=\tan(\frac{1}{2}\arcsin\epsilon)$. We may obtain a second way to express the Fourier series
\[
\frac{1}{r^2(t)} = \frac{1}{a^2} \sum_{n=-\infty}^\infty \mathrm e^{\mathrm in\omega t} \frac{1}{\sqrt{1-\epsilon^2}} \sum_{k=-\infty}^\infty J_k(n\epsilon)\varepsilon^{|n-k|}
.
\]
We can now pick out the Fourier coefficients of the light curve. They are, for $n\geq 1$,
\begin{equation}
f_n
= \frac{16}{3\pi\sqrt{1-\epsilon^2}} \sum_{k=-\infty}^\infty J_k(n\epsilon)\varepsilon^{|n-k|} \approx \frac{16\epsilon^n}{3\pi 2^n} \sum_{k=0}^n\frac{n^k}{k!}
.
\label{keplercoef}
\end{equation}
The second, approximate equality contains the lowest-order contributions in $\epsilon$.
The finite sum is the Taylor polynomial of degree $n$ for the function $\exp(n)$. Therefore the coefficients drop to zero exponentially if $\epsilon<2/\mathrm e$, as $f_n\sim\epsilon^n \mathrm e^n/2^n$.

Since Earth-like planets with a stable water-ice equilibrium will not likely occur for orbits that are too eccentric, we are mainly interested at low values of $\epsilon$. With the above equations it is possible to express the sine components as  linear combinations in $\epsilon$ of the cosine components for exact circular orbits.
\end{document}